\documentclass[%
 reprint,
showpacs,
 amsmath,amssymb,
 aps,
]{revtex4-1}

\usepackage{graphicx}
\usepackage{dcolumn}
\usepackage{bm}
\usepackage{bbm}
\usepackage{braket}
\usepackage{float}
\usepackage{chngcntr}
\usepackage{ulem}

\begin{document}


\title{Three-flavor collective neutrino conversions with multi-azimuthal-angle instability in an electron-capture supernova model}

\author{Masamichi Zaizen$^{1}$}
\email{mzaizen@astron.s.u-tokyo.ac.jp}
\author{Shunsaku Horiuchi$^{2}$}
\author{Tomoya Takiwaki$^{3}$}
\author{Kei Kotake$^{4,5}$}
\author{Takashi Yoshida$^{1}$}
\author{Hideyuki Umeda$^{1}$}
\author{John F. Cherry$^{6}$}

\affiliation{%
$^{1}$Department of Astronomy, Graduate School of Science, University of Tokyo, Tokyo 113-0033, Japan \\
$^{2}$Center for Neutrino Physics, Department of Physics, Virginia Tech, Blacksburg, VA 24061, USA \\
$^{3}$Division of Science, National Astronomical Observatory of Japan, 2-21-1 Osawa, Mitaka, Tokyo 181-8588, Japan \\
$^{4}$Department of Applied Physics, Fukuoka University, Nanakuma 8-19-1, Johnan, Fukuoka 814-0180, Japan\\
$^{5}$Research Insitute of Stellar Explosive Phenomena, Fukuoka University, Nanakuma 8-19-1, Johnan, Fukuoka 814-0180, Japan\\
$^{6}$Department of Physics, University of South Dakota, Vermillion, SD 57069, USA
}%

\date{\today}

\begin{abstract}
We investigate the multi-azimuthal angle (MAA) effect on collective neutrino oscillation by considering the three-dimensional neutrino momentum distribution in a realistic electron-capture supernova model with an $8.8 M_{\odot}$ progenitor.
We find that the MAA effect induces collective flavor conversions at epochs when it is completely suppressed under the axial-symmetric approximation.
This novel activity is switched on/off by the growth of the MAA instability and imprints additional time evolution in the expected neutrino event rate.
We validate our results by extending the linear stability analysis into the three-flavor scheme including mixing angles, and confirm that the onset of collective neutrino oscillation matches the steep growth of flavor instability.
We discuss how the MAA effect alters neutrino detection at Super-Kamiokande and DUNE. 
\end{abstract}


\maketitle


\section{Introduction}
The core collapse of a massive star leads to the formation of a proto-neutron star and emission of $\sim 10^{58}$ neutrinos.
The escaping neutrinos possess about $99\%$ of the gravitational energy released from the core collapse and, via neutrino-matter interactions, can power a shock wave.  If this process overpowers the matter accretion, the shock can eventually breakout of the photosphere and trigger a core-collapse supernova (CCSN) \cite{Langanke:2003,Mezzacappa:2005,Woosley:2005,Kotake:2006,Janka:2012,Janka:2017,Burrows:2013,Foglizzo:2015}.
In addition to the explosion mechanism, neutrinos have important roles in explosive nucleosynthesis.
Thus, the observation of neutrinos from nearby CCSNe helps us to confirm our understanding of the CCSN mechanisms and related physics \cite{Scholberg:2012,Horiuchi:2018}, e.g., provides information on the equation of state of nuclear matter from the spectral energy distribution.

Neutrino flavor conversions have a large influence on the contents of a detected neutrino signal \cite{Duan:2010,Mirizzi:2016,Chakraborty:2016}.
Since neutrinos mix in propagation, the original flavor information is changed by the time neutrinos are detected by terrestrial neutrino experiments.
In CCSNe, flavor mixing consists of three types of neutrino oscillations: vacuum oscillation, matter oscillation, and collective neutrino oscillation.
The first two are linear effects and well-understood by several neutrino experiments, and the only remaining unknown is the mass ordering problem.
In particular, the Mikheev-Smirnov-Wolfenstein (MSW) effect induced by background matter describes large flavor transitions at two typical electron densities \cite{Wolfenstein:1978,Mikheyev:1985}.
On the other hand, the third phenomenon is a complex non-linear effect induced by neutrino-neutrino coherent forward scattering interactions, and is still poorly understood.
The neutrino-neutrino interaction is important at high neutrino density and depends strongly on the neutrino trajectories and intersection angles.
The self-coupling interaction synchronizes flavor oscillation phases among different trajectories and causes collective neutrino oscillation.
The interaction system is a complicated seven-dimensional problem and often approximated by relaxing the complexity into simplified descriptions, e.g., the bulb model \cite{Duan:2006b}.
The bulb model requires many assumptions and symmetries on neutrino emission and background environments, especially axial symmetry in neutrino direction.
But it allows the computation of flavor conversions.

Collective neutrino oscillation is frequently initiated by a condition known as flavor instability and the behaviors of various flavor instabilities strongly depend on whether the neutrino mass ordering is normal or inverted.
Flavor evolution occurs due to the bimodal instability in the inverted mass ordering (IO), while the multi-zenith-angle (MZA) instability induces flavor conversion in the normal mass ordering (NO).
It is well known that collective neutrino oscillation presents simple spectral splits in the IO case \cite{Fogli:2007,Dasgupta:2008b}.
On the other hand, in the NO case, the MZA instability are often self-suppressed by neutrinos themselves especially in the early phase of CCSNe.
Since the neutrino mass ordering is still unsettled, many previous studies have investigated separately both mass ordering cases in order to verify the dependence on the mass ordering \cite{Esteban-Pretel:2007,Dasgupta:2009,Cherry:2010,Wu:2015,Sasaki:2017,Zaizen:2018,Sasaki:2020}.
However, recently global analyses including multiple neutrino experimental results suggest that the NO is favored over the IO at $\sim 3 \sigma$ level \cite{Capozzi:2017a,Capozzi:2018,deSalas:2018,Capozzi:2020a,deSalas:2020}.
While up-to-date neutrino oscillation data reduced the strong preference, the NO still remains preferred \cite{Kelly:2020,Esteban:2020}.

Linear stability analyses for collective neutrino oscillation have revealed that axial symmetry breaking produces a new instability, so-called the multi-azimuthal-angle (MAA) instability \cite{Raffelt:2013,Mirizzi:2013,Chakraborty:2014,Chakraborty:2014a}.
This instability could be enhanced even in NO under which the MZA instability is completely suppressed.
The axial-symmetry breaking corresponds to a departure from the bulb model approximation, moving from two-dimensional momentum space into a three-dimensional one.
This more precise model provides interesting differences between the MZA (bimodal) and MAA case, but makes the numerical investigation more challenging.

Previous numerical studies of MAA effects have presented significant flavor evolution in NO, under otherwise stable case (MZA) \cite{Mirizzi:2013,Chakraborty:2014,Chakraborty:2014a}.
However, many approximations have been used to reduce computational costs.
For example, previous works have employed non-realistic CCSN models for numerical simulations, using parametric supernova properties and neglecting the ordinary matter term by mimicking matter suppression \cite{Mirizzi:2013, Chakraborty:2014}.
When realistic models are used, monochromatic spectra are adopted instead of multi-energy distributions \cite{Chakraborty:2014a}.
Moreover, these investigations have assumed the two-flavor framework, not three-flavor. 
Under these assumptions, we cannot obtain accurate flavor conversions and estimates of neutrino detection.

In this paper, we perform the first-ever numerical study of the three-flavor collective neutrino oscillation considering three-dimensional momentum distribution in a realistic supernova model.
We focus on the NO case.
We also compare our numerical results with a linear stability analysis to confirm the validity of our numerical results. 
To this end, we extend two-flavor linear stability analyses in the literature to a three-flavor scheme including mixing angles.
Finally, we estimate the observed neutrino events at current and future neutrino detectors by using our obtained energy spectra.
The results are organized as follows.
In Sec.\ \ref{Sec2}, we introduce our numerical schemes for the hydrodynamical model and neutrino oscillation with MAA effects, and show the setup of three-flavor stability analysis including mixing angles.
In Sec.\ \ref{Sec3}, we present simulation results, compare them with those from linear analyses, and finally discuss the signal predictions from a Galactic CCSN event.
In Sec.\ \ref{Sec4}, we summarize our results and conclude.

\section{Methodology}
\label{Sec2}

\begin{figure}[t!]
	\centering
    \includegraphics[width=0.9\linewidth]{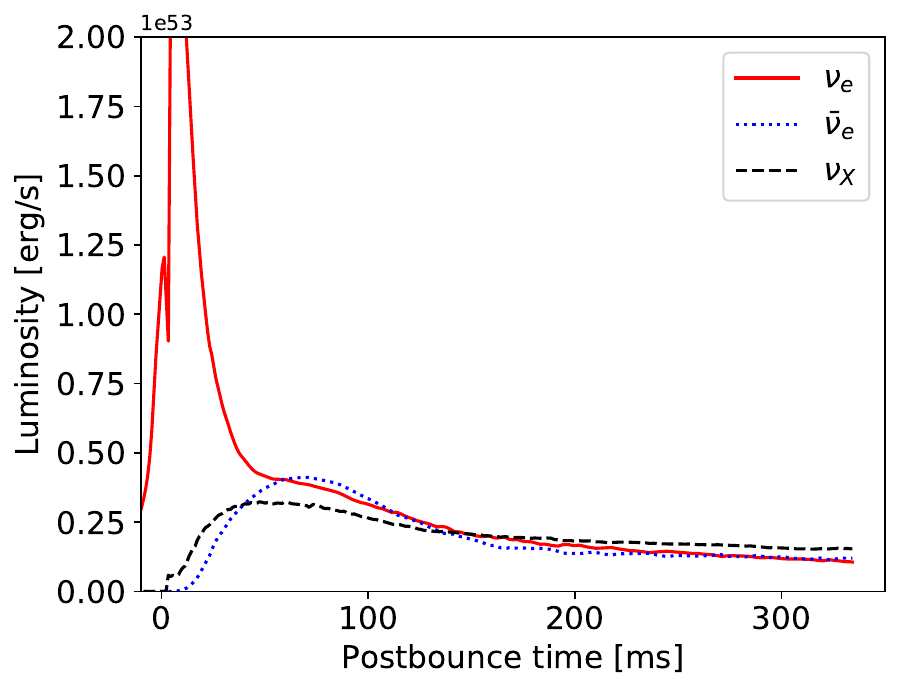}
    \includegraphics[width=0.9\linewidth]{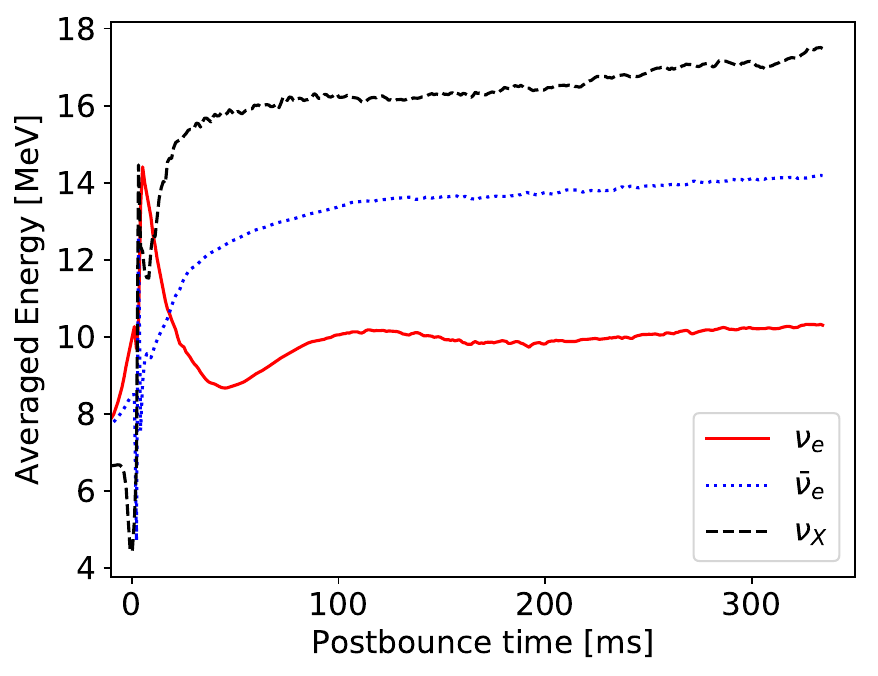}
    \includegraphics[width=0.9\linewidth]{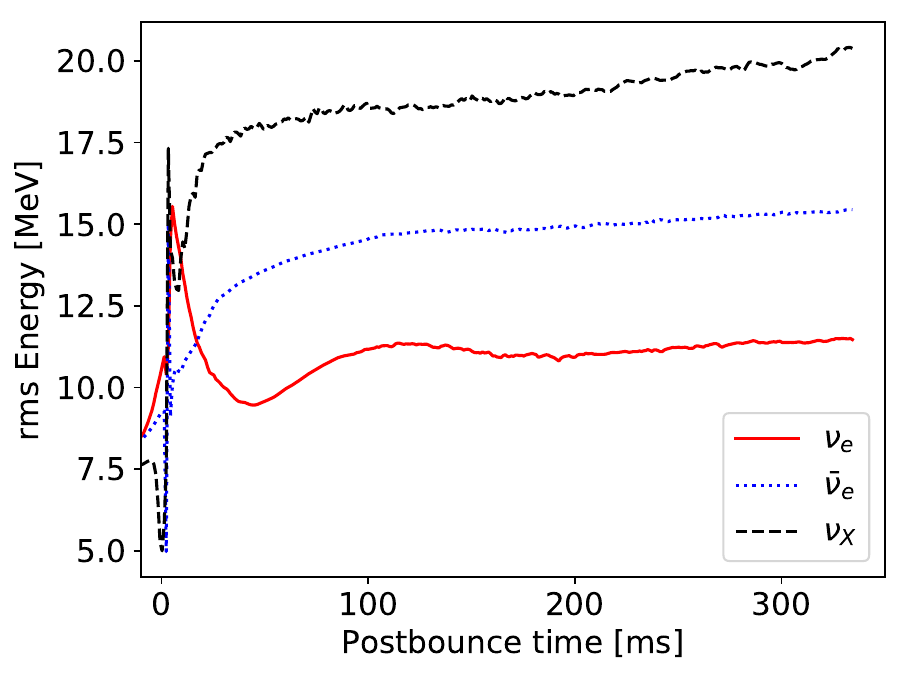}
	\caption{Time evolution of neutrino emission of the core collapse of an $8.8 M\odot$ progenitor.
	The top panel shows the neutrino luminosity, the center panel shows the mean energy, and the bottom panel shows the rms energy.
	Flavors $\nu_e$, $\bar{\nu}_e$, and $\nu_X$ ($=\nu_\mu,\bar{\nu}_\mu,\nu_\tau,\bar{\nu}_\tau$) are shown by red solid, blue dotted, and black dashed lines, respectively.
	}
	\label{fig:SN_data}
\end{figure}

\subsection{Supernova model}
We employ a two-dimensional electron-capture supernova (ECSN) model with an $8.8 M_{\odot}$ progenitor \citep{Tominaga:2013} (the setup of the envelope is same to that of Ref.~\citep{Kitaura:2006}).
As was done in Ref.~\cite{Zaizen:2020},
the hydrodynamic simulation has been performed by {\small 3DnSNe} code 
(see the references \citep{Nakamura:2019, OConnor:2018, Kotake:2018, Takiwaki:2016}
for recent applications).
The method for the hydrodynamic evolution is summarized in Ref.~\cite{Zaizen:2020}.
The two-dimensional simulation is computed on a spherical polar coordinate with spatial resolution of $(N_r, N_{\Theta})=(512,128)$. While the radial grid is logarithmically spaced and covers from 
0 to 5000 km, the polar grids covers from 0 to $\pi$ uniformly.

The equation of state used in the simulation is the Lattimer and Swesty with incompressibility of  $K=220\ {\rm MeV}$ \cite{Lattimer:1991}.
Although the code employs the relatively simple neutrino transport scheme of IDSA (Isotropic Diffusion Source Approximation) \cite{Liebendorfer:2009}, it nevertheless can provide consistent results on neutrino luminosities and average energies with more sophisticated schemes (see Ref.~\cite{OConnor:2018} for a detailed comparison).

Figure \ref{fig:SN_data} shows the time evolution of neutrino luminosity $L_{\nu}$, averaged neutrino energy $\langle E_{\nu}\rangle$, and rms energy $\sqrt{\langle E_{\nu}^2\rangle}$.
In these neutrino properties, $\nu_X$ means the non-electron type neutrinos $\nu_{\mu},\bar{\nu}_{\mu},\nu_{\tau}$, and $\bar{\nu}_{\tau}$.
We describe the initial neutrino spectra $\phi^i_{\nu}$ with a normalized Gamma distribution \cite{Keil:2003,Tamborra:2012,Tamborra:2014},
\begin{eqnarray}
    \phi^i_{\nu}(E) = \frac{E^{\xi}}{\Gamma_{\xi+1}}\left(\frac{\xi+1}{\langle E_{\nu}\rangle}\right)^{\xi+1}\exp\left[-(\xi+1)\frac{E}{\langle E_{\nu}\rangle}\right],
\end{eqnarray}
where the pinching parameter $\xi$ is defined by the averaged energy and rms energy as $\xi = \frac{\langle E_{\nu}^2\rangle-2\langle E_{\nu}\rangle^2}{\langle E_{\nu}\rangle^2 - \langle E_{\nu}^2\rangle}$. Time is measured after core bounce (i.e., post-bounce time) unless otherwise stated.

\subsection{Collective neutrino oscillation}
We extend the bulb model to three-dimensional momentum space by relaxing the assumptions of axial symmetry and investigate the effects of axial symmetry breaking on neutrino trajectories. 
Spatial spherical symmetry and azimuthal symmetry breaking can not be globally self-consistent.
When we extend the geometry to three-dimensional momentum space, the flavor evolution along the transverse direction to the neutrino propagation line should emerge under axial symmetry breaking.
Then, the flavor evolution is induced by both homogeneous and inhomogeneous modes \cite{Mangano:2014,Chakraborty:2016b}.
The inhomogeneity produces small-scale spatial variations but can only shift the flavor instability to more inner radius in the context of stability analysis.
Therefore, the traditional homogeneous mode still remains the most sensitive and dominant instability.
The small-scale modes can not break the multi-angle matter suppression when the largest-scale mode is completely suppressed.
However, the issue can be more complicated if we consider non-stationary modes.
In Refs.~\cite{Dasgupta:2015, Capozzi:2016}, the pulsating modes can compensate the matter-induced phase dispersion and pull the flavor instability up on the plane of matter density and radius.
Thereby, the inhomogeneous modes can become unstable even at inner radius at higher density.
The combination of the inhomogeneous and pulsating modes can have large influence on the flavor evolution.
However, as mentioned above, the possible impact of the transverse evolution is small unless we consider the non-stationarity simultaneously.

Here, we assume that the inhomogeneous variations always remain small and we ignore the transverse evolution as imposed in Refs.~\cite{Mirizzi:2013,Chakraborty:2014,Chakraborty:2014a}.
The assumption enables us to extend the bulb model into treatable description.
Also, we want to consider only the influence of the MAA instability, so we neglect the neutrino halo effect \cite{Cherry:2012,Cherry:2013,Cirigliano:2018,Zaizen:2020} and fast flavor conversions \cite{Sawyer:2005,Chakraborty:2016a,Izaguirre:2017,Dasgupta:2017,Abbar:2019,DelfanAzari:2020,Morinaga:2020,Glas:2020}.
In our work, we choose neutrino oscillation parameters as in Ref.~\cite{ParticleDataGroup:2018}: $\Delta m_{21}^2 = 7.37\times 10^{-5}\mathrm{~eV^2}$, $|\Delta m_{31}^2| = 2.56\times 10^{-3}\mathrm{~eV^2}$, $\sin^2\theta_{12}=0.297$, $\sin^2\theta_{13}=0.0214$, and CP-violation phase $\delta=0$.
We consider only the NO case $\Delta m_{31}^2 > 0$ because the MAA instability is more active and interesting for the NO case.
Also, recent experiments tend to favor it over IO \cite{Capozzi:2017a,Capozzi:2018,deSalas:2018,Capozzi:2020a,deSalas:2020,Kelly:2020,Esteban:2020}.

The flavor evolution is described as the evolution for a density matrix $\rho_{\nu}$.
The equation of motion along the radial direction in steady state is
\begin{eqnarray}
    i \partial_r \rho_{\nu} &&= \left[H^{+}_{E,u,\varphi},\rho_{\nu}\right] \\
    \label{diffevol}
    i \partial_r \overline{\rho}_{\nu} &&= \left[H^{-}_{E,u,\varphi},\overline{\rho}_{\nu}\right]
\end{eqnarray}
\begin{widetext}
\begin{equation}
    H^{\pm}_{E,u,\varphi} = \frac{1}{v_{r,u}}\left(\pm U\frac{M^2}{2E}U^{\dagger}+\sqrt{2}G_{\mathrm{F}}n_e L\right) + \frac{\sqrt{2}G_{\mathrm{F}}}{4\pi r^2}\int\frac{\mathrm{d}E^{\prime}\mathrm{d}u^{\prime}\mathrm{d}\varphi^{\prime}}{2\pi} \left(\frac{1-v_{r,u}v_{r,u^{\prime}}-\boldsymbol{\beta}\cdot\boldsymbol{\beta}^{\prime}}{v_{r,u}v_{r,u^{\prime}}}\right) \left(\rho_{\nu}^{\prime}-\overline{\rho}_{\nu}^{\prime}\right),
\end{equation}
\end{widetext}
where $U$ is the Pontecorvo-Maki-Nakagawa-Sakata matrix \cite{Maki:1962}, $M^2$ is a neutrino mass square matrix, $n_e$ is the electron number density, and $L$ is $\mathrm{diag(1,0,0)}$.
The radial velocity $v_{r,u}$ with an angular mode $u$ is defined as
\begin{eqnarray}
    v_{r,u} &&= \cos\theta =  \sqrt{1-u\frac{R_{\nu}^2}{r^2}} \\
    u &&= \sin^2\theta_{R},
\end{eqnarray}
where $\theta$ is an intersection angle and $\theta_{R}$ is an emission angle relative to the radial direction on the surface of an emission source $R_{\nu}$, which is fixed at $30\mathrm{~km}$.
Finally, $\boldsymbol{\beta}$ is the transverse velocity and the azimuthal-angle term 
\begin{eqnarray}
    \boldsymbol{\beta}\cdot\boldsymbol{\beta}^{\prime} &&= \sin\theta\sin\theta^{\prime}\cos(\varphi-\varphi^{\prime}) \notag\\
    &&= \sqrt{u u^{\prime}}\frac{R_{\nu}^2}{r^2}\cos(\varphi-\varphi^{\prime})
\end{eqnarray}
breaks the traditional axial symmetry \cite{Raffelt:2013}.
If the initial condition $\rho_{\nu}^{i}$ possesses any axial perturbations, the non-axial symmetry can grow and lead to flavor conversions.
If the initial neutrino flux is axial symmetric, the azimuthal angle term is equal to zero due to the periodic integral of the cosine and this MAA Hamiltonian $H_{E,u,\varphi}$ is identical to $H_{E,u}$ in the traditional bulb model.
Note that non-zero numerical errors in the periodic integral $\int\mathrm{d}\varphi$ for axial-symmetric flux act like initial perturbations and such artificial seeds can grow via the MAA instability even if we do not provide any instability seed in the azimuthal angle distribution.

We use a resolution of $N_E=200$ between $[0,60]\mathrm{~MeV}$, $N_u=2048$ between $[0,1]$, and $N_{\varphi}=64$ between $[0,2\pi]$ when binning the  three-dimensional momentum space $(E,u,\varphi)$.
We have checked that numerical results with $N_{\varphi}=64$ are identical to those with $N_{\varphi}=128$.
The azimuthal-angle resolution is less than the polar-angle one.
The different polar-angular modes travel along the different distance and the matter-induced phase dispersion requires the fine polar-angle resolution.
On the other hand, the travel path does not change on the each azimuthal-angular mode and the matter-induced potential does not affect the azimuthal-angle resolution unless we employ the multi-dimensional matter background.

We describe the flavor difference with polarization vectors $\mathbf{P}$ and $\overline{\mathbf{P}}$ in solving the equation of motion.
Here, we adopt the rotated frame of $(e-x-y)$ flavor state instead of $(e-\mu-\tau)$ flavor state \cite{Dasgupta:2008b,Zaizen:2018}.
This rotated frame enables us to understand three-flavor conversions easily.
The polarization vector reflects independent components of the density matrix and we can transform Eq.~(\ref{diffevol}) from matrix differential equations including complex values to vector differential equations composed only of real values as
\begin{eqnarray}
\rho_{\nu} &&= \frac{1}{3}I_3 +\frac{1
}{2}\mathbf{P}\cdot\mathbf{\Lambda} \\
\partial_r\mathbf{P}(E,u) &&= \mathbf{H}_{E,u,\varphi} \times \mathbf{P}(E,u,\varphi),
\end{eqnarray}
where $\mathbf{\Lambda}$ is a vector of the Gell-Mann matrices.
This three-flavor formalism is based on Ref.~\cite{Dasgupta:2008b,Zaizen:2018}.
Then, $P_3$ and $P_8$ correspond to the diagonal terms of the density matrix $\rho_{\nu}$ and initial conditions are written as
\begin{eqnarray}
&&P^i_3(E,u,\varphi) = f_{\nu_e}(E,u,\varphi) - f_{\nu_x}(E,u,\varphi) \\
&&P^i_8(E,u,\varphi) = \frac{f_{\nu_e}(E,u,\varphi) +f_{\nu_x}(E,u,\varphi) -2f_{\nu_y}(E,u,\varphi)}{\sqrt{3}}. \notag \\
\end{eqnarray}
Initial axial perturbation included in neutrino flux $f_{\nu_{\alpha}}$ leads to the MAA instability via the azimuthal-angle term.
Here, we assume cosinusoidal axial angular perturbation as initial conditions
\begin{equation}
    f_{\nu_{\alpha}}(E,u) \to f_{\nu_{\alpha}}^{\prime}(E,u,\varphi) = f_{\nu_{\alpha}}(E,u)\left(1+\varepsilon\cos\varphi\right),
\end{equation}
where we set perturbation with $\varepsilon \sim 1\%$.
This produces non-zero azimuthal-angle integration and gives instability seed into the MAA Hamiltonian $H_{E,u,\varphi}$.
\\

\subsection{Linear analysis}
Here, we introduce our linear analysis framework, which we will use to check our numerical results (see Sec.~\ref{Sec3b}).
Many previous works have adopted a simple two-flavor system ignoring mixing angles in order to probe the possibility of flavor oscillation, but these results cannot faithfully reproduce three-flavor effects.
Especially, recent works suggested that non-small mixing angle $\theta_{13}$ leads to the new instability associated with $\Delta m^2_{21}$ in NO \cite{Doring:2019,Sasaki:2020}.
In the three-flavor framework, the adopted values of mixing angles become more important.
So we develop a three-flavor stability analysis with the format including mixing angles $\theta_{12}$ and $\theta_{13}$.

In order to see the evolution of the off-diagonal terms of the density matrix, we can decompose it into
\begin{eqnarray}
    \rho_{\nu} &&= \frac{\mathrm{Tr}(\rho_{\nu})}{3}\mathbbm{1} + \frac{f_{\nu_e}-f_{\nu_x}}{3}\begin{pmatrix} s_1 & S_1 & 0 \\
    S_1^* & -s_1 & 0 \\
    0 & 0 & 0
    \end{pmatrix} \notag \\
    &&~~~~~~~~~~~~~~+ \frac{f_{\nu_e}-f_{\nu_y}}{3}\begin{pmatrix} s_2 & 0 & S_2 \\
    0 & 0 & 0 \\
    S_2^* & 0 & -s_2
    \end{pmatrix} \notag \\
    &&~~~~~~~~~~~~~~+ \frac{f_{\nu_x}-f_{\nu_y}}{3}\begin{pmatrix} 
    0 & 0 & 0 \\
    0 & s_3 & S_3 \\
    0 & S_3^* & -s_3
    \end{pmatrix} \\
    &&= \frac{\mathrm{Tr}(\rho_{\nu})}{3}\mathbbm{1} + \sum^{3}_{j=1}g_j\mathbb{S}_j,
\end{eqnarray}
where complex $S_j$ and real $s_j$ describes flavor coherence and flavor conversion, respectively.
If the employed supernova model gives $f_{\nu_x} = f_{\nu_y}$, the third term, $j=3$, is canceled.
In this situation, the $e-x$ and $e-y$ sectors are decoupled in the linear regime \cite{Chakraborty:2020}.

In a fast-rotating frame \cite{Banerjee:2011}, the vacuum matrix is simply expressed as
\begin{eqnarray}
    H_{\mathrm{vac}} \to \Braket{H_{\mathrm{vac}}} &&= \frac{1}{2E}\begin{pmatrix}
    0 & 0 & 0 \\
    0 & \Delta m^2_{21}& 0 \\
    0 & 0 & \Delta m^2_{31}
    \end{pmatrix}\notag \\
    &&= \mathrm{diag}(0,\omega_L,\omega_H).
\end{eqnarray}
Subscripts $H/L$ are in conjunction with the notation of the MSW resonances and express the $e-y$ and $e-x$ sectors, respectively.
To linear order, we can assume $\left|S_j\right|\ll 1$ and $s_j=1$, and we obtain the linearized equation for $S_j$ as
\begin{eqnarray}
    i v_{r,u}\partial_r S_j &&= (\mu_j+\lambda_j-\omega_j)S_j \notag \\
    && - \sqrt{2}G_{\mathrm{F}}\int\mathrm{d}\Gamma^{\prime}(1-\mathbf{v}\cdot\mathbf{v}^{\prime})(g_j^{\prime}S_j^{\prime}-\bar{g}_j^{\prime}\bar{S}_j^{\prime}),
\end{eqnarray}
where
\begin{eqnarray}
    &&\mu_j = \sqrt{2}G_{\mathrm{F}}\int\mathrm{d}\Gamma^{\prime}(1-\mathbf{v}\cdot\mathbf{v}^{\prime})(g_j^{\prime}-\bar{g}_j^{\prime}) \\
    &&\lambda_1 = \lambda_2 = \lambda = \sqrt{2}G_{\mathrm{F}}n_e,~ \lambda_3 = 0 \\
    &&\omega_1 = \omega_L,~ \omega_2 = \omega_H,~ \omega_3 = \omega_H - \omega_L.
\end{eqnarray}
Especially, in the identical non-electron type case $g_3=0$ and isotropic emission case, the self-interaction potential $\mu_j$ is simply
\begin{eqnarray}
    &&\mu_1 = \mu_2 = \epsilon\mu,~ \mu_3 = 0, \\
    &&\epsilon\mu = \sqrt{2}G_{\mathrm{F}}\frac{\Phi_{\nu_e}-\Phi_{\bar{\nu}_e}}{4\pi r^2},
\end{eqnarray}
where $\Phi_{\nu_{\alpha}} = L_{\nu_{\alpha}}/\langle E_{\nu_{\alpha}}\rangle$ is a neutrino flux of $\alpha$ flavor and $\epsilon$ is a flavor asymmetry parameter \cite{Esteban-Pretel:2007} defined as
\begin{equation}
    \epsilon = \frac{\Phi_{\nu_e}-\Phi_{\bar{\nu}_e}}{\Phi_{\bar{\nu}_e}-\Phi_{\bar{\nu}_x}}.
\end{equation}
In the following, we re-define the energy distribution $g_j$ normalized by $(\Phi_{\bar{\nu}_e}-\Phi_{\bar{\nu}_x})$ as $\int\mathrm{d}\Gamma g_j = \int^{\infty}_{-\infty}\mathrm{d}\omega_j \int^{1}_{0}\mathrm{d}u \int^{2\pi}_{0}\frac{\mathrm{d}\varphi}{2\pi}g_j = \epsilon$.
Under the large-distance approximation $r\gg R_{\nu}$,

\begin{equation}
    v_{r,u}^{-1} = \left(1-u\frac{R_{\nu}^2}{r^2}\right)^{-1/2} \simeq 1 + u\frac{R_{\nu}^2}{2r^2}.
\end{equation}
Therefore, the linearized equation is given by
\begin{eqnarray}
    &&i\partial_r S_j = \left(u\bar{\lambda}_j-\omega_j\right)S_j \notag \\
    &&~~- \mu^*\int\mathrm{d}\Gamma^{\prime}\left[u+u^{\prime}-2\sqrt{uu^{\prime}}\cos(\varphi-\varphi^{\prime})\right]g_j^{\prime}S_j^{\prime}. \label{eq:S-eq}
\end{eqnarray}
Here, we adopt the effective matter potential $\bar{\lambda}_j$ by using the multi-angle potential $\mu^*$ and $\lambda^*$ as follows:
\begin{equation}
    \bar{\lambda}_j = \mu_j^*+\lambda_j^* = (\mu_j+\lambda_j)\frac{R_{\nu}^2}{2r^2}.
\end{equation}

We can write the solutions as $S_j = Q_j \mathrm{e}^{-i \Omega_j r}$ in the Fourier modes.
The non-trivial solutions for eigenvectors $Q_j$ are
\begin{eqnarray}
    D_j(\Omega_j)
    &&\equiv (I_{j,1}-1)^2 - I_{j,0}I_{j,2} = 0 ~~~(\mathrm{for~ MZA}) \\
    &&\equiv I_{j,1}+1 = 0 ~~~(\mathrm{for~ MAA}),
\end{eqnarray}
where
\begin{eqnarray}
    I_{j,n} = \mu^*\int\mathrm{d}\omega_j\mathrm{d}u\frac{u^n g_j}{u\bar{\lambda}_j-\omega_j-\Omega_j}. \label{eq:Ijn}
\end{eqnarray}
The flavor conversions could occur under conditions satisfying the exponential growth, $\mathrm{Im}(\Omega_j) \equiv \kappa_j > 0$.

However, the stability analysis in this fast-rotating frame does not provide the new instability associated with $\Delta m^2_{21}$ in NO because it completely ignores mixing angles.
We therefore follow the approach in Ref.~\cite{Doring:2019} and diagonalize the Hamiltonian of linear effects:
\begin{eqnarray}
    &&H_{\mathrm{diag}}^{\prime} \notag \\
    &&\approx \begin{pmatrix}
    \lambda+\omega_H(s_{13}^2+\eta c_{13}^2 s_{12}^2) & 0 & 0 \\
    0 & \omega_H \eta c_{12}^2 & 0 \\
    0 & 0 & \omega_H(c_{13}^2+\eta s_{12}^2 s_{13}^2)
    \end{pmatrix}, \notag \\
\end{eqnarray}
where $\eta = \omega_L/\omega_H$ is the mass-squared difference ratio, $s_{ij}=\sin\theta_{ij}$, and $c_{ij} = \cos\theta_{ij}$.
Comparing this $H_{\mathrm{diag}}^{\prime}$ with $H_{\mathrm{diag}}$ in the fast-rotating frame, we find simple substituting relations
\begin{eqnarray}
    \lambda~ &&\to \lambda+\omega_H(s_{13}^2+\eta c_{13}^2 s_{12}^2) \\
    \omega_L &&\to \omega_H \eta c_{12}^2 \\
    \omega_H &&\to \omega_H(c_{13}^2+\eta s_{12}^2 s_{13}^2).
\end{eqnarray}
We substitute these relations into Eq.~\eqref{eq:Ijn} considering the deformation into Eq.~\eqref{eq:S-eq},
\begin{eqnarray}
    &&I_{1,n} = \mu^*\int\mathrm{d}\omega_H\mathrm{d}u\frac{u^n g_1}{u\bar{\lambda}-\omega_H L_{\mathrm{mix}}-\Omega_1} \\
    &&I_{2,n} = \mu^*\int\mathrm{d}\omega_H\mathrm{d}u\frac{u^n g_2}{u\bar{\lambda}-\omega_H H_{\mathrm{mix}}-\Omega_2},
    \label{eq:DR_eq}
\end{eqnarray}
where
\begin{eqnarray}
    &&L_{\mathrm{mix}} = \eta( c_{12}^2-s_{12}^2 c_{13}^2)-s_{13}^2 \\
    &&H_{\mathrm{mix}} = \cos 2\theta_{13}(1-\eta s_{12}^2).
\end{eqnarray}
By using these expanded stability analysis, we can investigate the flavor conversions within three-flavor framework.
\\

\section{Results}
\label{Sec3}
In the following, we first present our numerical results on collective flavor conversions including the MAA effects in Sec.~\ref{Sec3a}.
Then, we compare them with our linear stability analyses in Sec.~\ref{Sec3b}.
We finally discuss how the MAA instability influences the neutrino detection rates at Super-Kamiokande and DUNE in Sec.~\ref{Sec3c}.

\subsection{Flavor conversion}
\label{Sec3a}
\begin{figure}[t]
	\centering
    \includegraphics[width=0.9\linewidth]{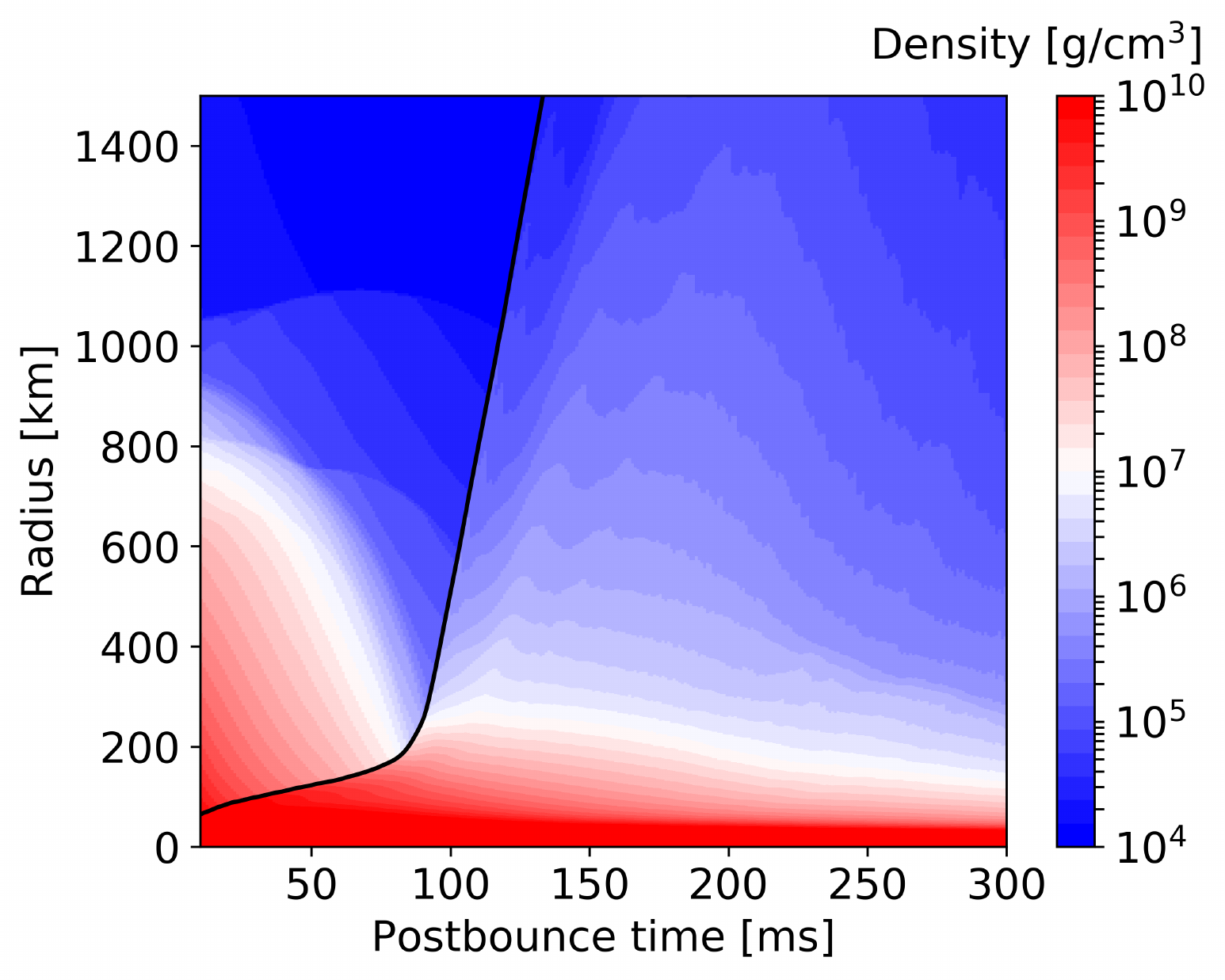}
    \includegraphics[width=0.9\linewidth]{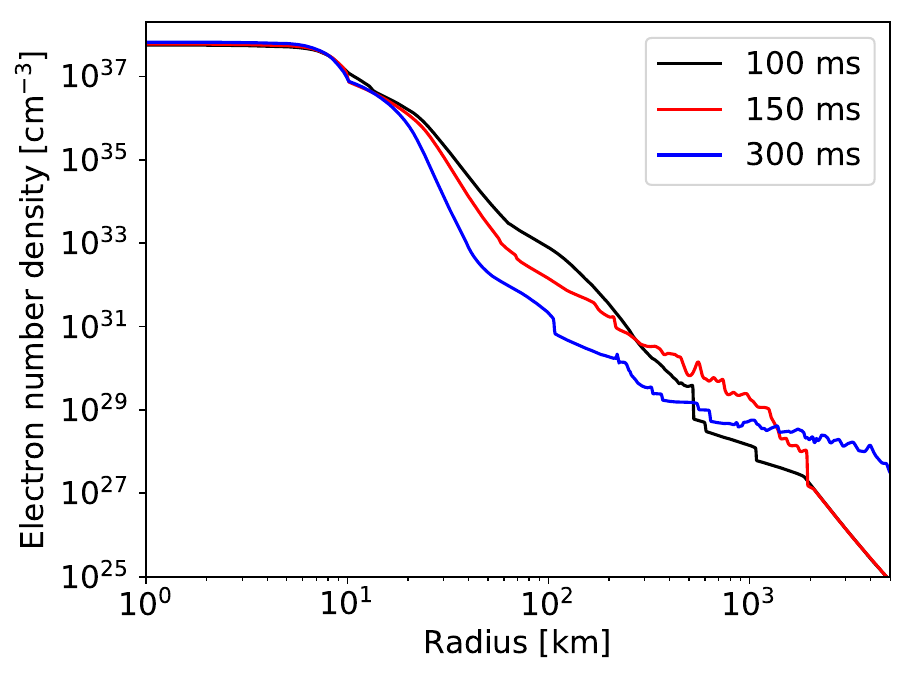}
	\caption{Top: Black line is the time evolution of the averaged shock radius. Colored contours represent the time evolution of the averaged density profile.
	Bottom: Electron number density profile $n_e$ along the north pole at postbounce time $t_{\rm pb} = 100, 150$, and $300\mathrm{~ms}$.
	}
	\label{fig:density_prof}
\end{figure}

The top panel of Figure \ref{fig:density_prof} shows the time evolution of the averaged shock radius (black curve) and the matter density (color contours).
The (neutrino-driven) shock expansion  can be seen at around postbounce time $t_{\mathrm{pb}}=100\mathrm{~ms}$, and the accreting matter is blown out.
Thereby, the matter density increases until around $200\mathrm{~ms}$ at radii where collective neutrino oscillation can be induced.
The growth of the shock front has a large influence on the activity of collective flavor conversion, and we can divide the behaviors into four phases.
As we explain below, these are the suppression phase, window phase, re-suppression phase, and revival phase.

The bottom panel of Figure \ref{fig:density_prof} shows the electron number density along the northern pole at the representative snapshots of each phase, $100$, $150$, and $300 \mathrm{~ms}$.
During the early epoch of SNe, collective neutrino oscillation is suppressed by the excess of electron neutrinos due to the neutronization burst and heavy accreting matter; this is the suppression phase which corresponds to a timescale of $t_{\mathrm{pb}} < 100\mathrm{~ms}$.
After that, shock propagation creates a window of time when collective effects can overcome matter-induced phase dispersion; this is the window phase lasting from $100\mathrm{~ms} \leq t_{\mathrm{pb}} < 150\mathrm{~ms}$.
As time goes on, the ejected shocked material later increases the matter density and the matter effects again suppress the collective flavor conversions; this is the re-suppression phase and spans the times $150\mathrm{~ms} \leq t_{\mathrm{pb}} < 250\mathrm{~ms}$.
Finally, the self-induced effects revive during the cooling epoch as the matter density decreases; this is the revival phase beginning at $t_{\mathrm{pb}} = 250\mathrm{~ms}$.

\begin{figure}[t]
	\centering
    \includegraphics[width=0.9\linewidth]{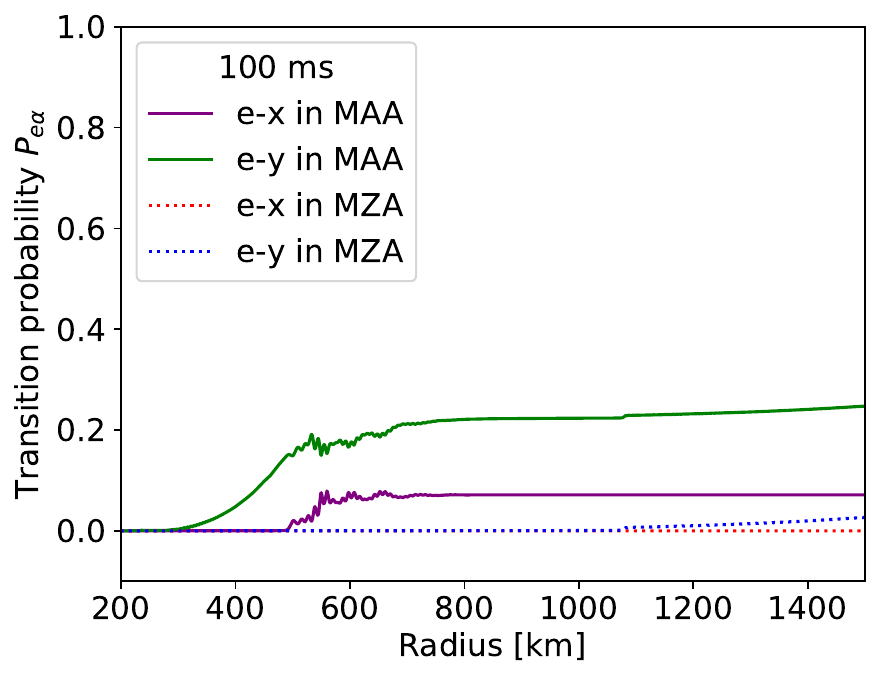}
    \includegraphics[width=0.9\linewidth]{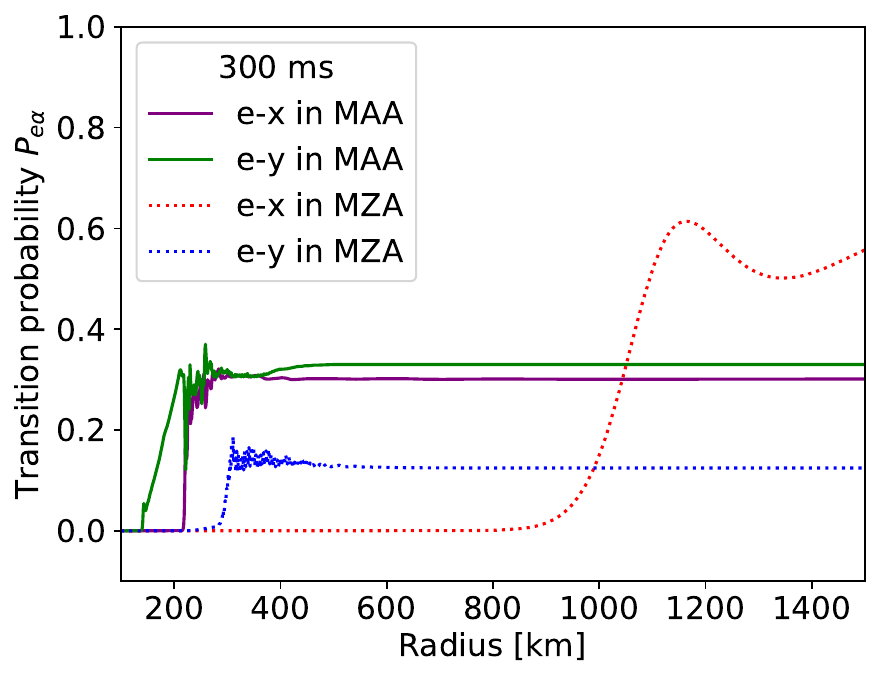}
	\caption{Radial evolution of the transition probability of electron neutrinos into non-electron types at $100$ ms (top panel) and $300\mathrm{~ms}$ (bottom panel).
	Solid lines correspond to the MAA case and dotted lines correspond to the MZA case.
	At $100\mathrm{~ms}$, the MZA instability is completely suppressed, while the MAA effects lead to flavor conversions. 
	At $300\mathrm{~ms}$, flavor conversions are seen in both MZA and MAA cases, especially beyond $1000 \mathrm{~km}$ where oscillation appears in the MZA case in the $e-x$ sector.
	Also, transition probabilities are almost identical to $1/3$ in the MAA case.
	}
	\label{fig:surv}
\end{figure}

Figure \ref{fig:surv} shows the radial evolution of the transition probability of electron neutrinos averaged over energy and angular distributions at $100$ and $300\mathrm{~ms}$.
Collective neutrino oscillation is completely suppressed at $150\mathrm{~ms}$ in both cases and we do not show it in this figure.
Here, we divide it into two types of transition probability in $e-x$ sector and $e-y$ sector.
These two sectors are associated with solar mass-squared difference $\Delta m^2_{21}$ and atmospheric mass-squared difference $\Delta m^2_{31}$, respectively.
Note that the two-flavor frameworks adopted by previous works do not show the three-flavor effects and can only provide flavor conversion in the $e-y$ sector.

At $100\mathrm{~ms}$ (top panel of Figure \ref{fig:surv}), collective neutrino oscillation only occurs in the MAA case (solid lines); the MZA case (dotted lines) is completely suppressed.
The different behavior stems from the large flavor asymmetry $\epsilon$ at this snapshot and it gives stable flavor mode for MZA case in NO \cite{Esteban-Pretel:2007}.
In the MAA case (solid lines), the $e-y$ sector starts flavor mixing from $250\mathrm{~km}$ and the $e-x$ sector from around $500\mathrm{~km}$.
This difference in the onset radius results from the balance between the vacuum term and the collective term.
Collective neutrino oscillation arising in our calculation is sometimes called ``slow'' mode and is induced on scales of the vacuum frequency $\omega_{H/L}$, different from fast flavor conversions \cite{Sawyer:2005}.
The oscillation frequency $\mu$ is slower as the neutrino density decreases and the growth rate becomes large at the radius satisfying $\mu\sim\omega$.
Consequently, the onset for the $e-x$ sector associated with $\Delta m^2_{21}$ is more delayed by the mass-squared difference ratio $\eta$ compared to the $e-y$ sector.
This is a simple understanding and actually the onset radius also depends on mixing angles.
We will discuss this dependence in Sec.\ref{Sec3b}.
Although $e-x$ flavor conversions occur, the transition probability is less than $10\%$.
The $e-y$ flavor conversions remain dominant and hence it is akin to a two-flavor system even in the MAA case at $100\mathrm{~ms}$.
Transition probability in the $e-y$ sector gradually appears to increase beyond $1000\mathrm{~km}$ in both the MAA and MZA cases, but these are induced by the H-resonance of matter oscillation, not by self-interactions.

At $300\mathrm{~ms}$ (bottom panel of Figure \ref{fig:surv}), the situation is largely different and flavor conversions in both the MAA and MZA cases occur.
$e-y$ conversions in the MZA case (dotted blue line) starts from $200\mathrm{~km}$ and grows most around $300\mathrm{~km}$.
Especially, the $e-x$ sector in the MZA case (dotted red line) shows large flavor mixing beyond around $1000\mathrm{~km}$.
The matter density does not still reach the region satisfying $\lambda \sim \omega_H$ and this peculiar flavor conversion is not related to the H-resonance as observed at $100\mathrm{~ms}$.
This snapshot is a case with a flux ordering commonly called multiple crossings and this case has been investigated in Refs.~\cite{Dasgupta:2009,Friedland:2010,Dasgupta:2010,Mirizzi:2011}.
Multiple crossings provide three-flavor effects leading to $e-x$ mixing and this behavior can happen in both IO and NO.
On the other hand, the onset radius in the MAA case is much smaller than in the MZA cases for both $e-x$ and $e-y$ sectors.
In the MAA case, $e-y$ conversions start from $130\mathrm{~km}$ and $e-x$ conversions from $220\mathrm{~km}$.
$e-x$ conversions in the MAA case is quicker compared to the MZA case.
The reason will be discussed in Sec.~\ref{Sec3b}.
And the transition probabilities for ordinary neutrinos of each flavor lead to almost $1/3$ and hence the neutrino ensemble appears to reach flavor equilibrium.
This multi-angle decoherence induced by the MAA instability was reported by Ref.~\cite{Mirizzi:2013}.
Neutrinos do not reach flavor equilibrium at $100\mathrm{~ms}$ when flavor asymmetry $\epsilon$ is large.
This fact is consistent with previous works within the two-flavor framework \cite{Mirizzi:2013}.
\begin{figure}[t]
	\centering
    \includegraphics[width=0.9\linewidth]{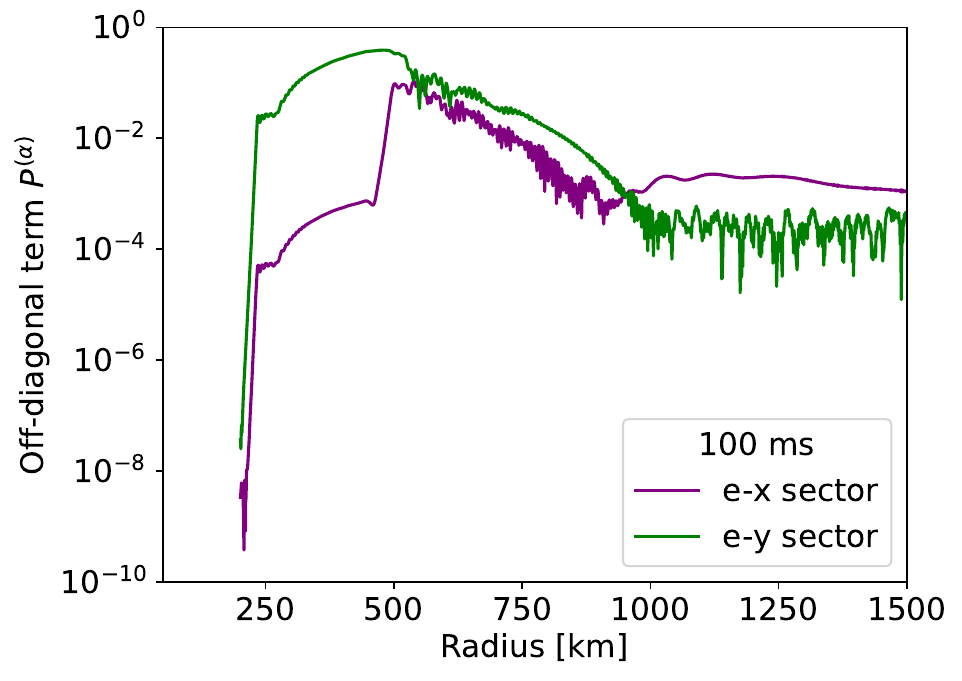}
    \includegraphics[width=0.9\linewidth]{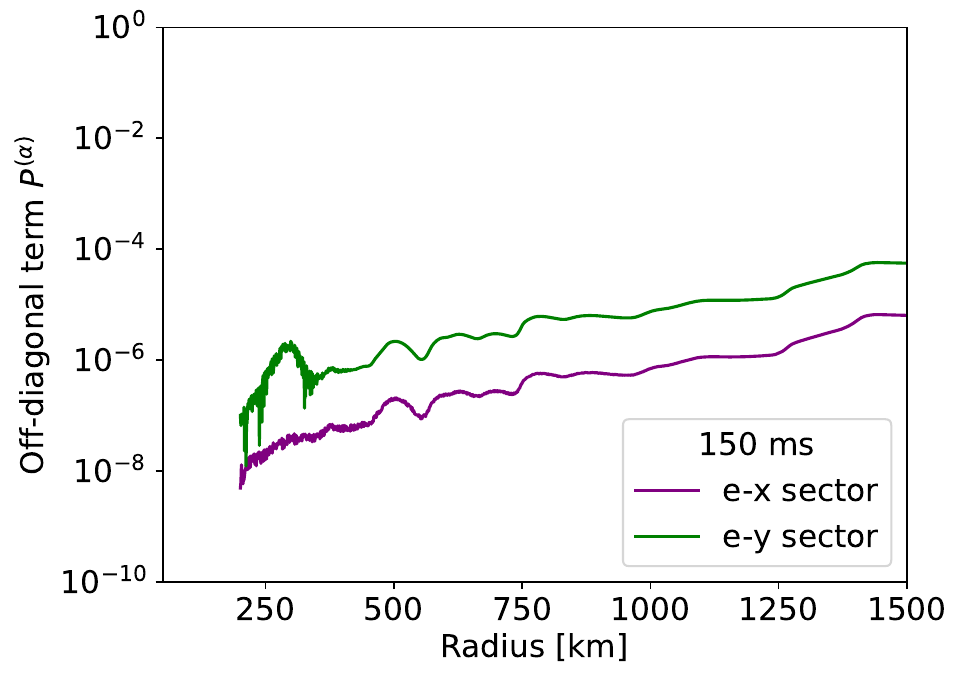}
    \includegraphics[width=0.9\linewidth]{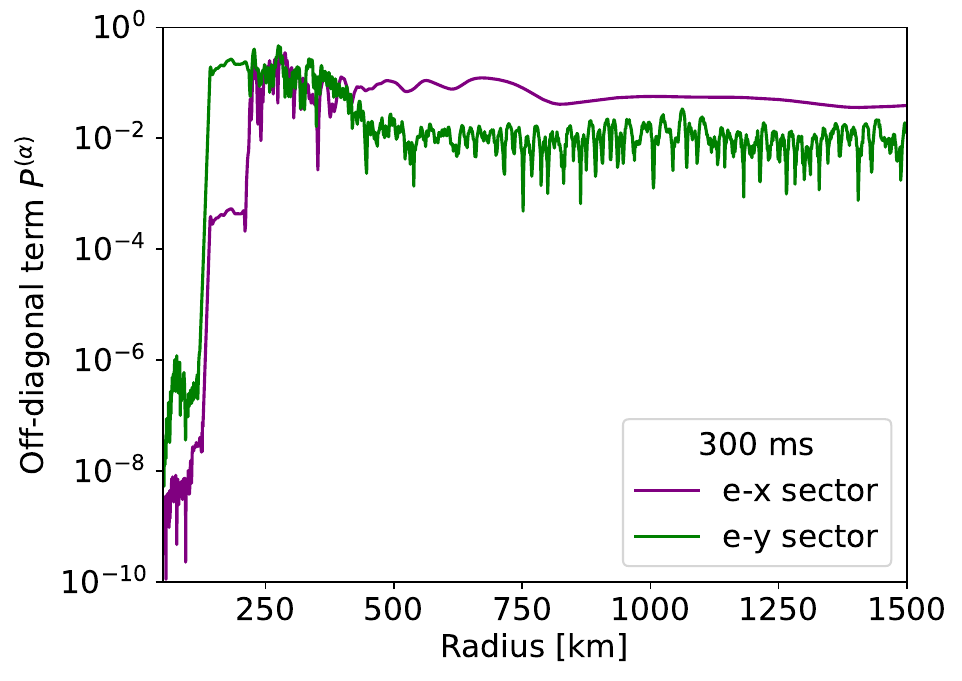}
	\caption{Radial evolution of the dipole term in $\cos\varphi$ of off-diagonal terms at $100$ ms (top), $150$ ms (middle), and $300\mathrm{~ms}$ (bottom).
	Purple lines correspond to the $e-x$ sector and green lines to the $e-y$ sector.
	At $100$ ms and $300\mathrm{~ms}$, flavor instability steeply grows, while the growth is suppressed at $150\mathrm{~ms}$.
	}
	\label{fig:dipole}
\end{figure}

\begin{figure*}[t]
	\centering
	\begin{minipage}{0.45\hsize}
    \includegraphics[width=0.9\linewidth]{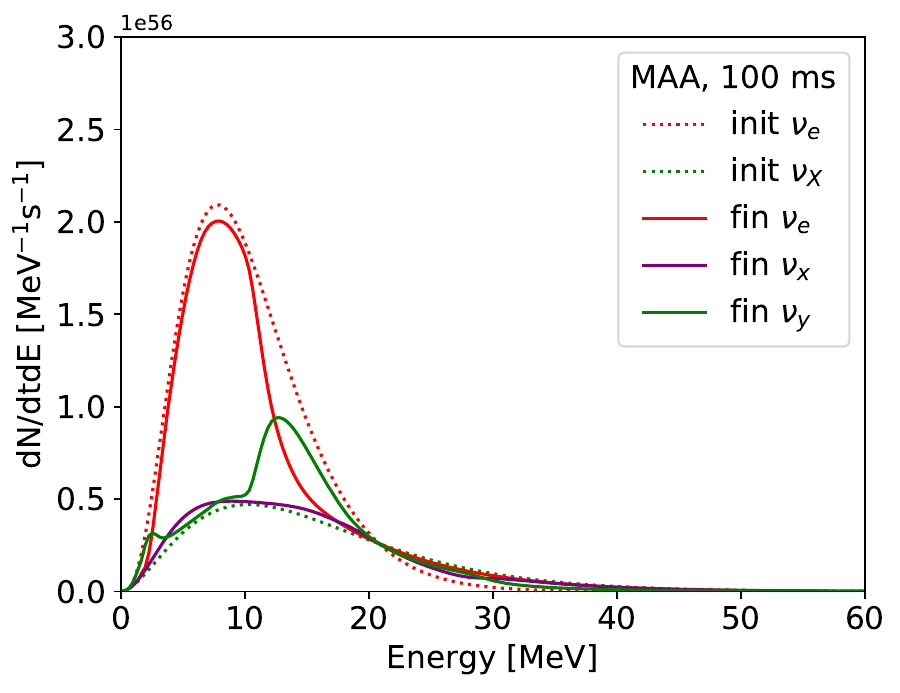}
    \includegraphics[width=0.9\linewidth]{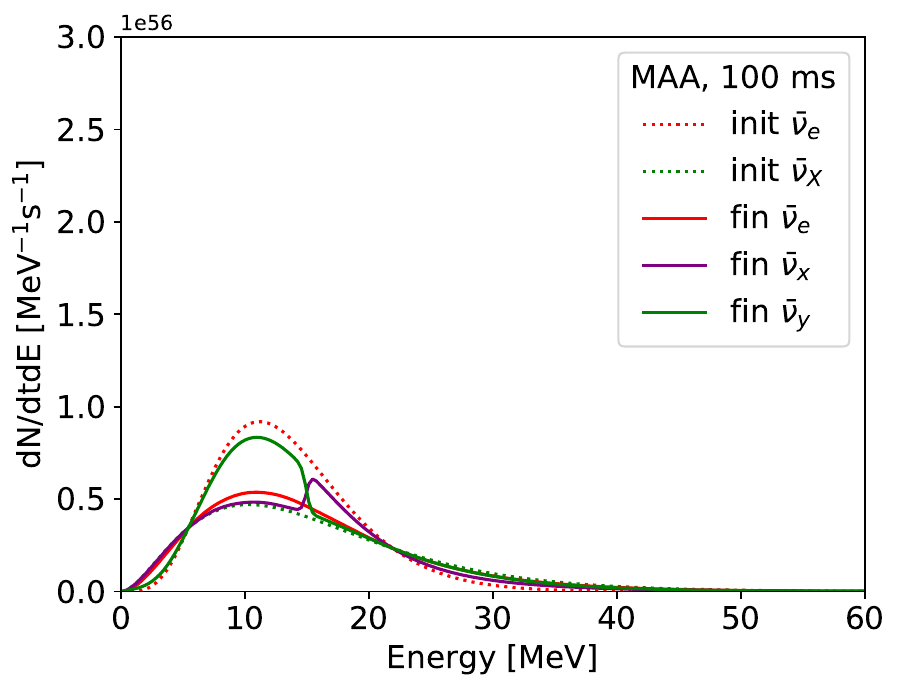}
    \end{minipage}
    \begin{minipage}{0.45\hsize}
    \includegraphics[width=0.9\linewidth]{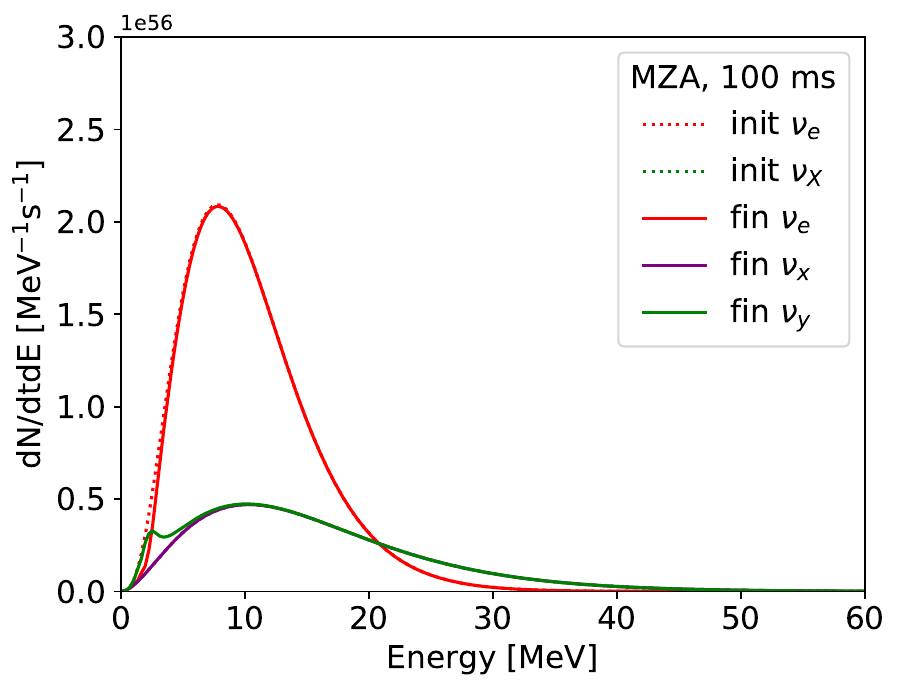}
    \includegraphics[width=0.9\linewidth]{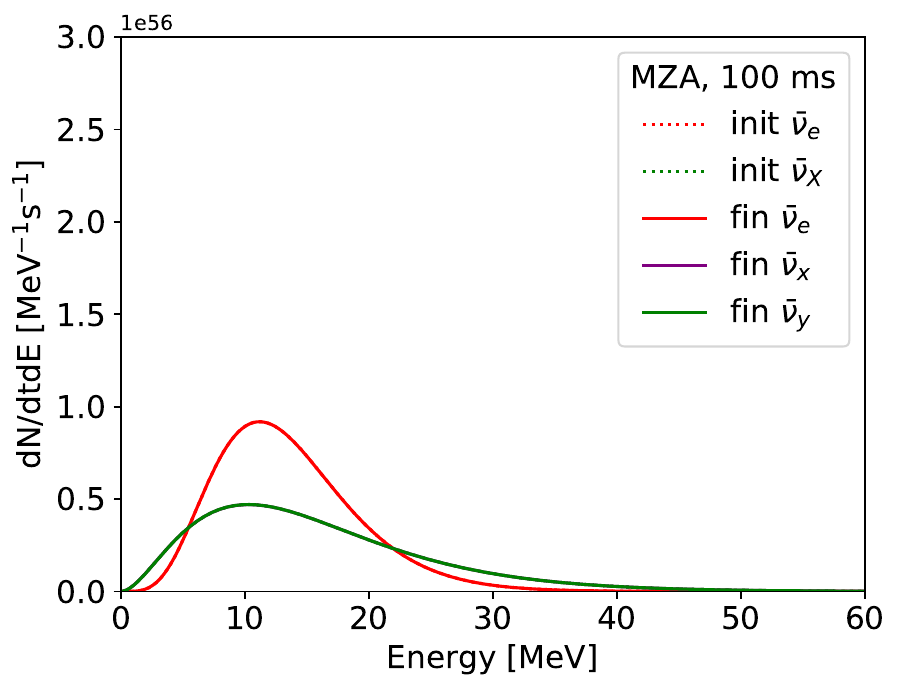}
    \end{minipage}
	\caption{Neutrino and anti-neutrino spectra at $1500\mathrm{~km}$ at $100\mathrm{~ms}$.
	Left panels are the MAA case and right panels are the MZA case.
	Top panels are neutrino spectra and bottom panels are the anti-neutrino sector.
	The dotted lines are for original spectra and the solid ones are for final spectra after collective neutrino oscillation. In the MAA case, spectral splits appear, while flavor conversions are completely suppressed in the MZA case.
	Low-energy transitions in neutrino spectra occur due to the H-resonance.
	}
	\label{fig:sp_100}
\end{figure*}

Figure \ref{fig:dipole} shows the radial evolution of the dipole term in $\cos\varphi$ of the off-diagonal term of the density matrix $\rho_{\nu}$ at $100$, $150$, and $300\mathrm{~ms}$.
This value indicates the growth of azimuthal-angle instability \cite{Mirizzi:2013} and is defined as
\begin{eqnarray}
    &&P^{(x)} = \sqrt{(P^{(1)})^2+(P^{(2)})^2} \\
    &&P^{(y)} = \sqrt{(P^{(4)})^2+(P^{(5)})^2},
\end{eqnarray}
where
\begin{eqnarray}
    P^{(m)} = \int\frac{\mathrm{d}E\mathrm{d}u\mathrm{d}\varphi}{2\pi}P^{(m)}_{E,u,\varphi}\cos\varphi.
\end{eqnarray}
These $P^{(1,2,4,5)}$ correspond to real and imaginary components of the off-diagonal terms in the density matrix $\rho_{\nu}$ and they are initially identical to zero.
Flavor conversions driven by the MAA instability occur at a radius where this dipole term approaches of order unity.
If other instability dominantly leads to flavor conversions, this dipole term does not need to sufficiently grow.
Comparing these off-diagonal terms with transition probabilities in the MAA case, steep growing features are directly linked to the onset radius of flavor conversions.
In other words, flavor conversions at $100$ and $300\mathrm{~ms}$ are derived by the MAA instability.
Also, we find that this flavor instability cannot grow sufficiently due to the dense background matter at $150\mathrm{~ms}$ in spite of given perturbation seeds.

Figure \ref{fig:sp_100} shows neutrino and anti-neutrino spectra at $1500\mathrm{~km}$ at $100\mathrm{~ms}$.
Left panels are the MAA case and right panels are the MZA case.
Top panels are neutrino spectra and bottom panels are the anti-neutrino sector.
At this time snapshot, flavor conversions are completely suppressed in the MZA case and propagating neutrinos only experience the H-resonance beyond $1000\mathrm{~km}$.
The H-resonance provides almost complete $e-y$ conversion for the neutrino sector in the NO case, and we find spectral swaps only in the low-energy region less than $5\mathrm{~MeV}$ because neutrinos do not reach high energy region yet.
In the MAA case, spectral splits in the $e-y$ sector occur beyond $10\mathrm{~MeV}$ and they are tiny in the $e-x$ sector.
In the anti-neutrino sector, the final electron anti-neutrino spectrum is almost converted to the original non-electron type spectrum.
And $x-y$ conversion is seen above $15\mathrm{~MeV}$.
These are three-flavor effects and arise due to the non-linear evolution.
These swap features resemble spectral splits induced by the bimodal instability in IO \cite{Sasaki:2020}.
This is due to the large flavor asymmetry $\epsilon>1$ and we observe the quasi-single angle behavior \cite{Esteban-Pretel:2007,Mirizzi:2013}.
Bump features in the $\nu_y$ spectrum are transferred into electron neutrinos through the H-resonance at larger radius.

\begin{figure*}[t]
	\centering
	\begin{minipage}{0.45\hsize}
    \includegraphics[width=0.9\linewidth]{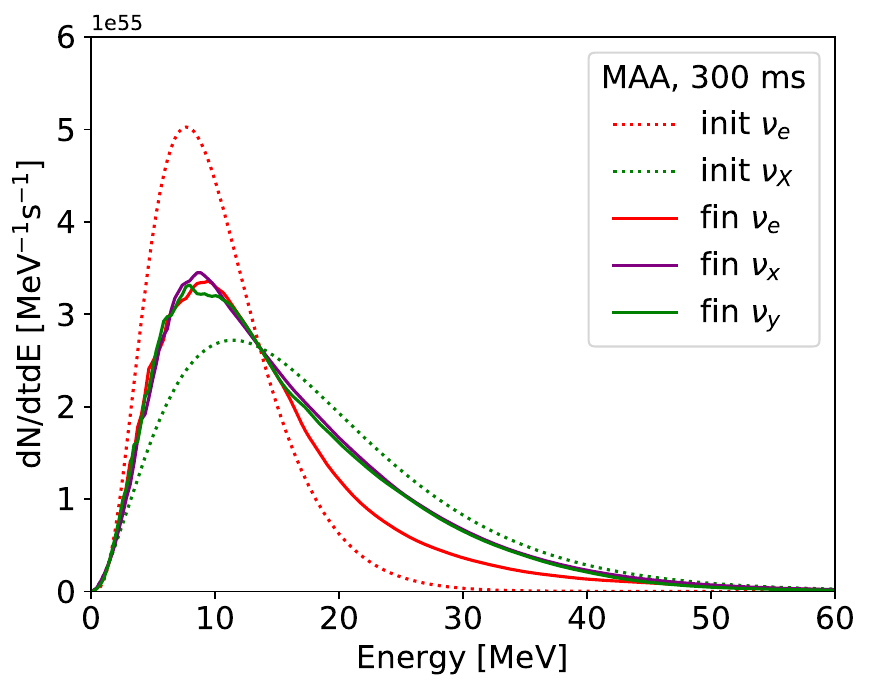}
    \includegraphics[width=0.9\linewidth]{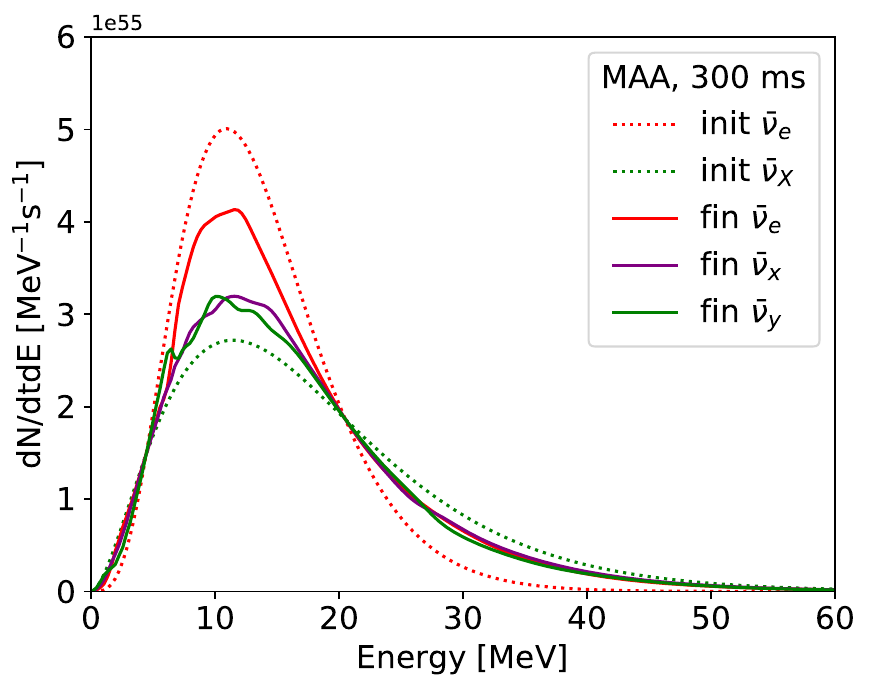}
    \end{minipage}
    \begin{minipage}{0.45\hsize}
    \includegraphics[width=0.9\linewidth]{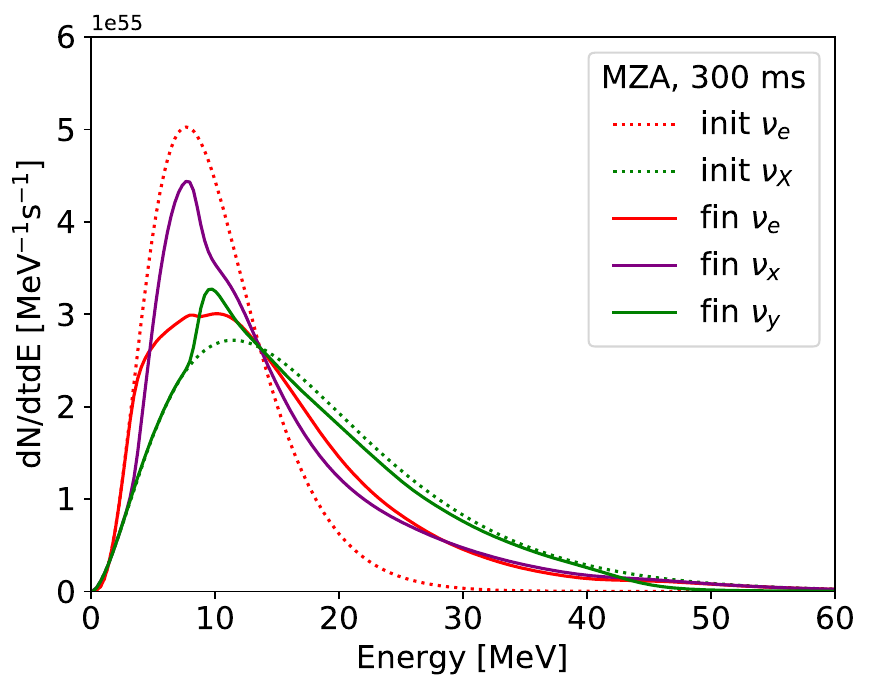}
    \includegraphics[width=0.9\linewidth]{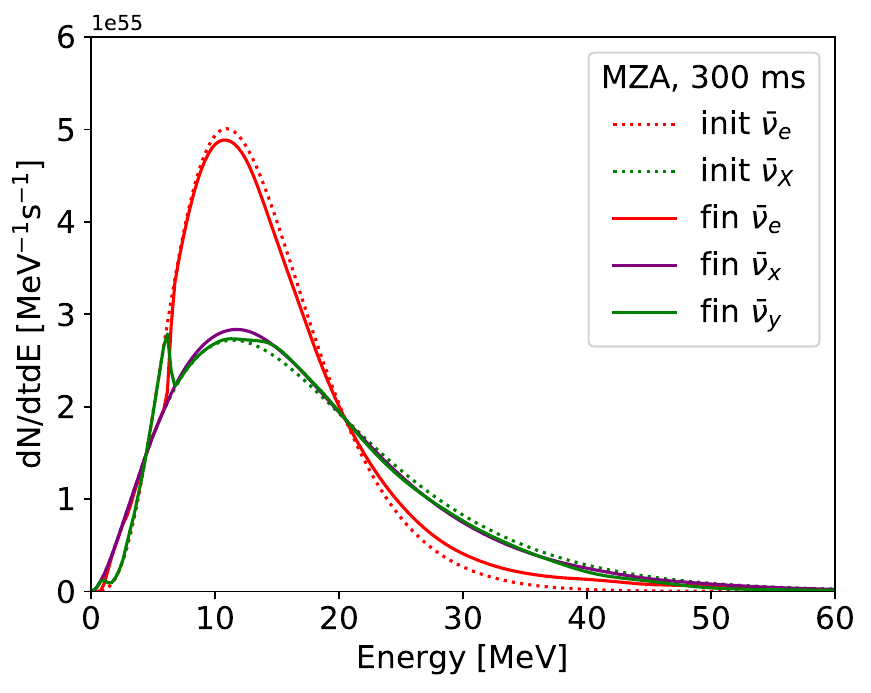}
    \end{minipage}
	\caption{The same as Fig. \ref{fig:sp_100}, but for $300\mathrm{~ms}$.
	Neutrino spectra of three flavors are almost identical and there are no splits in the MAA case.
	The same features are observed for anti-neutrinos.
    In the MZA case, split features in the neutrino sector are shifted by $e-x$ conversion subsequent to $e-y$ conversion.
    In the anti-neutrino case, $e-x$ conversion is tiny and a small low-energy splits occurs in the $e-y$ sector only.
	}
	\label{fig:sp_300}
\end{figure*}

\begin{figure}[b]
	\centering
    \includegraphics[width=0.9\linewidth]{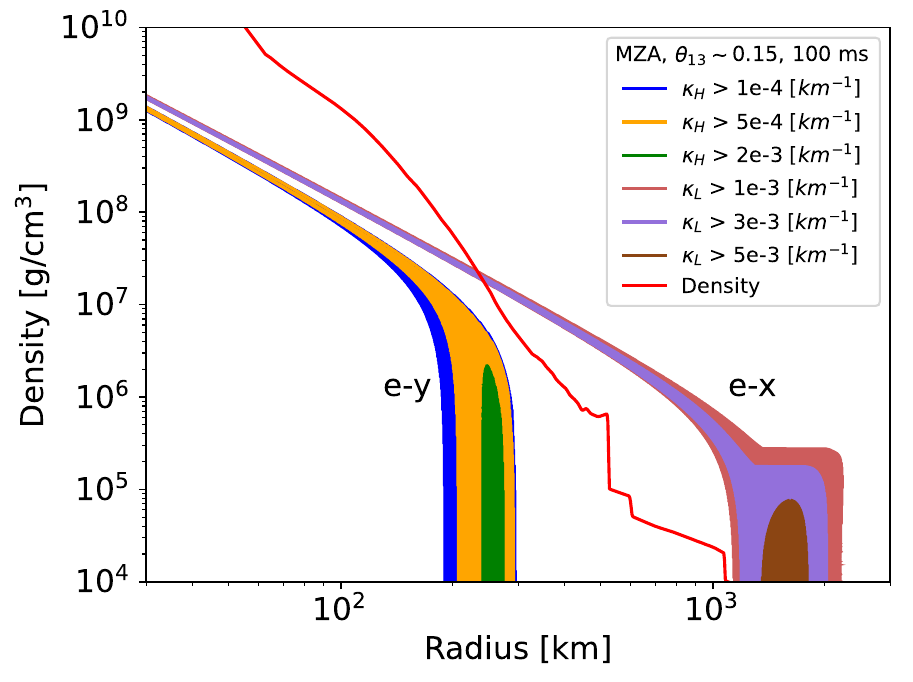}
	\centering
    \includegraphics[width=0.9\linewidth]{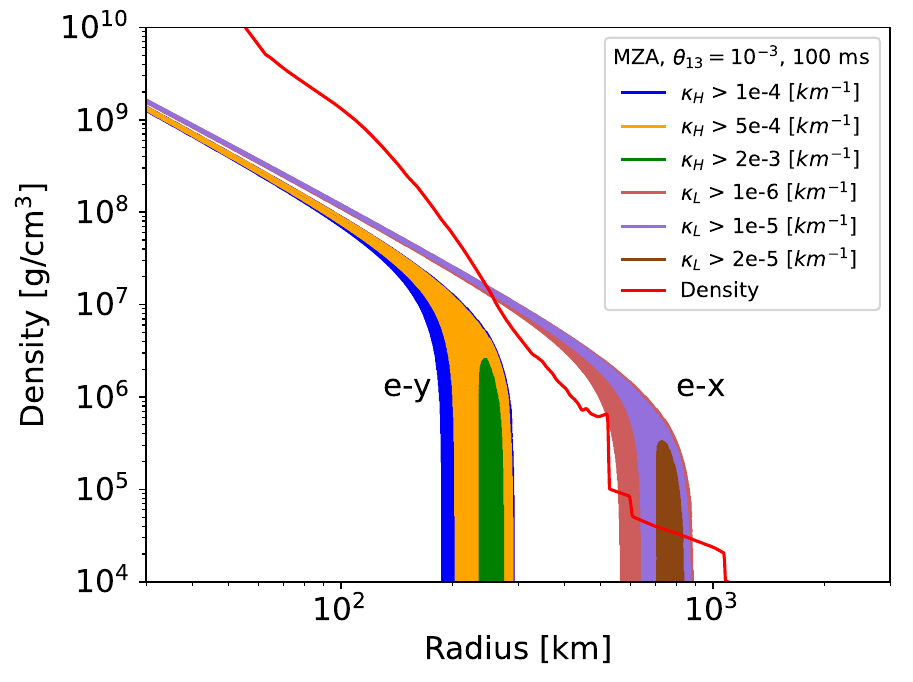}
	\caption{Contour map of the growth rate $\kappa_{H/L}$ in the MZA case at $100\mathrm{~ms}$.
	Top panel is for ordinary mixing angle case and bottom is for small mixing angle case, $\theta_{13} = 10^{-3}$.
	The regions for the $e-y$ and $e-x$ sectors are labeled.
	The red line shows the density profile.
	Flavor instability can grow if the density profile intersects the patched regions.
	}
	\label{fig:foot_100_MZA}
\end{figure}
\begin{figure}[b]
	\centering
    \includegraphics[width=0.9\linewidth]{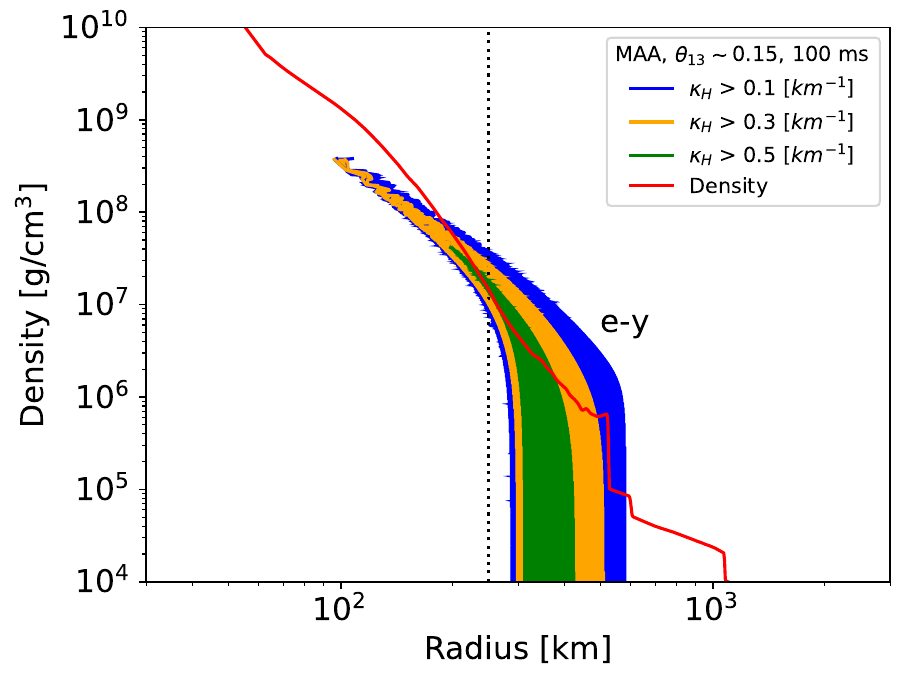}
    \includegraphics[width=0.9\linewidth]{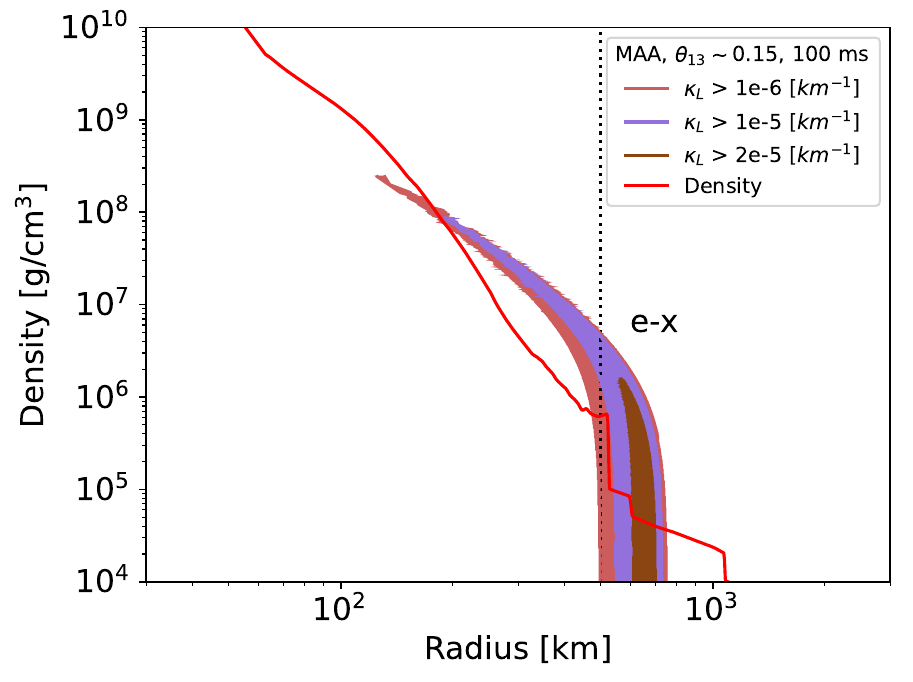}
    \includegraphics[width=0.9\linewidth]{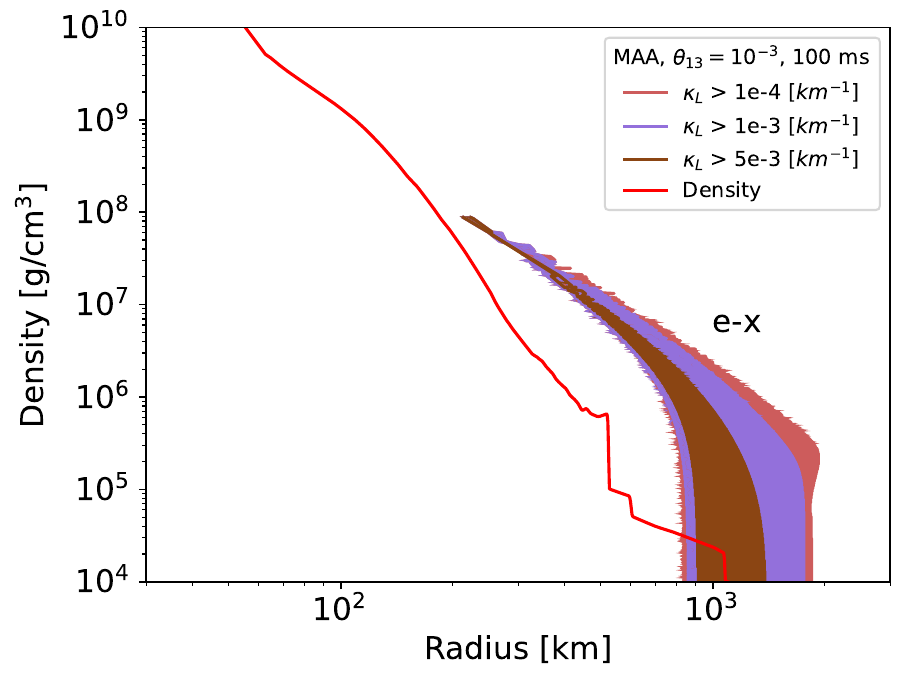}
	\caption{The same as Fig.~\ref{fig:foot_100_MZA}, but for the MAA case.
	Here, $e-y$ (top) and $e-x$ (middle and bottom) sectors are separately plotted.
	We do not show the plot in $e-y$ sector in the small mixing angle case because it is almost identical as the ordinary case.
	The vertical dotted line indicates the onset radius of collective neutrino oscillations found by our numerical treatment in each sector.
	}
	\label{fig:foot_100_MAA}
\end{figure}

Figure \ref{fig:sp_300} shows the neutrino and anti-neutrino spectra at $300\mathrm{~ms}$.
At this time snapshot, flavor conversions occur in both the MAA and MZA cases.
Averaged transition probabilities for each flavors in the neutrino sector are almost $1/3$ in the MAA case and the final spectra are actually identical below a crossing energy of $15\mathrm{~MeV}$.
This behavior is also seen in the anti-neutrino sector and three-flavor states reach the same number density above $20\mathrm{~MeV}$.
Hence, the neutrino ensemble obtains a partial flavor equilibrium through the MAA effects.
And MSW resonances do not have large effects on the final spectra under flavor equilibrium.
On the other hand, flavor equilibrium is not established in the MZA case.
Spectral splits arising in the $e-y$ sector in neutrino spectra at around $300\mathrm{~km}$ are transferred into the $\nu_x$ spectrum through large radius oscillations in the $e-x$ sector.
Swap structures escape into the non-electron neutrinos but return through subsequent MSW resonances.
\\

\subsection{Results of stability analysis}
\label{Sec3b}
In this section, we discuss our linear stability analyses to interpret our numerical results. 
In particular, we focus on three-flavor effects on the MAA instability because previous works have discussed only within the two-flavor framework.
We show the growth rate maps of the MZA and MAA instability at $100$ and $300\mathrm{~ms}$ obtained by solving Eq.~\eqref{eq:DR_eq}.

Figure \ref{fig:foot_100_MZA} shows contours of growth rates $\kappa_{H/L}$ in the MZA case at $100\mathrm{~ms}$.
In the top panel, the density profile (solid red line) passes through between the instability regions in both the $e-x$ and $e-y$ sectors, and this situation is consistent with our numerical results that flavor conversions do not happen.
The density profile does not deeply invade into the fast growth regions and instead goes across the narrow region of $e-x$ sector growth.
This is not enough for flavor instability to grow because the growth rate immediately becomes weaker again.

The behaviors of three-flavor effects in the MZA case depend on the value of $\theta_{13}$ as mentioned in Ref.~\cite{Doring:2019}.
Adopting a small or zero value, the stability analysis presents different growth rate especially in the $e-x$ sector.
In order to understand the effects of the mixing angle, we discuss the effects within the three-flavor framework adopting a small mixing angle $\theta_{13} = 10^{-3}$, as was done in Refs.~\cite{Dasgupta:2008b, Dasgupta:2010, Mirizzi:2011}.
This is shown in the bottom panel of Fig.~\ref{fig:foot_100_MZA}.
Comparing these two mixing angle cases, the instability region in the $e-x$ sector is shifted into smaller radii.
In addition, the growth rate is significantly smaller.
If we ignore the mixing angles and solve Eq.~\eqref{eq:Ijn} and not Eq.~\eqref{eq:DR_eq}, the instability region in the $e-x$ sector is shifted to smaller radii.
Therefore, flavor evolution can be suppressed because of the small growth rate even if the density profile intersects the instability region.
Even in the ordinary case, adopting experimental values, the growth rate in the $e-x$ sector is still smaller, but it sufficiently satisfies $\kappa_L r > 1$ and is larger than in the $e-y$ sector.
There is a possibility that collective flavor conversion can occur only in the $e-x$ sector, depending on the model investigated as mentioned in Ref. \cite{Doring:2019}.
Also, we observe that the instability region in the $e-y$ sector depends very weakly on the value of $\theta_{13}$.
So long as we discuss the possibility of flavor evolution within the two-flavor framework, the inclusion of $\theta_{13}$ does not have any influence.
However, it is necessary to take mixing angles into account on the linear stability analysis if performing three-flavor calculations as in this work.

Figure \ref{fig:foot_100_MAA} shows the instability contour maps in the MAA case at $100\mathrm{~ms}$.
We divide the $e-x$ and $e-y$ sectors into two separate panels (top and middle) because the instability regions partly overlap.
We show the onset radius of collective neutrino oscillation found by our numerical treatment in each sector as a vertical dotted line.
This indicator clearly corresponds to the intersection point of the density profile (red solid) with the instability regions.
It is easy to see it even in comparison with the radial evolution of the dipole mode of off-diagonal terms in Figure \ref{fig:dipole}.
Dipole modes in the off-diagonal terms steeply evolve at a radius where the propagating neutrinos obtain large growth rates inside the instability regions.
Compared to the growth rate in the $e-y$ sector, that in the $e-x$ sector is much smaller and does not satisfy the condition $\kappa_L r > 1$.
Nevertheless, $e-x$ flavor conversion appears in our numerical results.
This is because in the linear regime, we only calculate the linearized equations by considering the two sectors are completely decoupled.
But in the numerical treatment, the system is non-linear and the rapid evolution of the $e-y$ sector kicks instability seeds in the $e-x$ sector and the growth is more rapid compared to that predicted \cite{Dasgupta:2010}.
Such behaviors also appear in the radial evolution of the dipole mode of the off-diagonal terms.
Indeed, the growth of instability in the $e-x$ sector starts to develop only after the $e-y$ sector evolves.

The dependence of the instability regions on mixing angles in the $e-x$ sector has been discussed in the MZA case.
Since it is likely to appear in the MAA case, we similarly perform the stability analysis.
Interestingly, the bottom panel adopting small mixing angle $\theta_{13} = 10^{-3}$ shows larger growth rates that are also shifted to larger radii compared to the non-small mixing angle case shown in the middle panel.
This is contrary to the MZA case, where the non-small mixing angle $\theta_{13}$ enhances the instability associated with $\Delta m^2_{21}$.
This difference in behavior stems from the different oscillation mechanisms between the MAA and MZA.
According to Ref.~\cite{Sasaki:2020}, the sign of the Bloch vector $\boldsymbol{B}$ of the vacuum Hamiltonian affects whether flavor mixing becomes unstable in the bulb model or not, 
\begin{eqnarray}
    B_{e-y}~ &&\sim -\frac{\Delta m^2_{31}}{2E}\cos 2\theta_{13} \label{eq:Bysign}\\
    B_{e-x}~ &&\sim -\frac{\Delta m^2_{21}}{2E}\left(\cos 2\theta_{12}-\cos^2\theta_{12}\sin^2\theta_{13}\right) \notag\\
    &&~~~~+\frac{\Delta m^2_{31}}{2E}\sin^2\theta_{13}. \label{eq:Bxsign}
\end{eqnarray}
In the IO case, Eq.~\eqref{eq:Bysign} is always positive due to the sign of $\Delta m^2_{31}$ and the positive sign is preferable for unstable flavor instability.
In NO, Eq.~\eqref{eq:Bysign} is always negative and $e-y$ conversions are stable in the MZA case.
On the other hand, the sign of Eq.~\eqref{eq:Bxsign} depends on the value of mixing angle $\theta_{13}$ and the selection of a sufficiently large value changes the sign from negative to positive.
In that case, flavor stability in the $e-x$ sector becomes unstable and obtains faster growth rates compared to the small mixing angle case.
On the other hand, $e-y$ conversions occur in the MAA case though the MZA case is potentially stable.
It can be considered that the MAA effect reverses the meanings of the sign and the negative is preferable for unstable instability.
This is consistent with the behaviors that the dependence of the growth rate in the $e-x$ sector on the mixing angle $\theta_{13}$ is contrary to the MZA case.
Also, the intersection point of the density profile in the $e-x$ sector occurs at approximately $800\mathrm{~km}$ and obviously does not match the onset radius $\sim 500\mathrm{~km}$ in Figure \ref{fig:surv}, though the position in the $e-y$ sector changes little.
As with the MZA case, the instability region is more shifted if we ignore mixing angles.
Thus, the linear stability analysis scheme ignoring mixing angles or using different values can incorrectly indicate the possibility of flavor conversions or lack thereof within three-flavor framework.

\begin{figure}[h]
	\centering
    \includegraphics[width=0.9\linewidth]{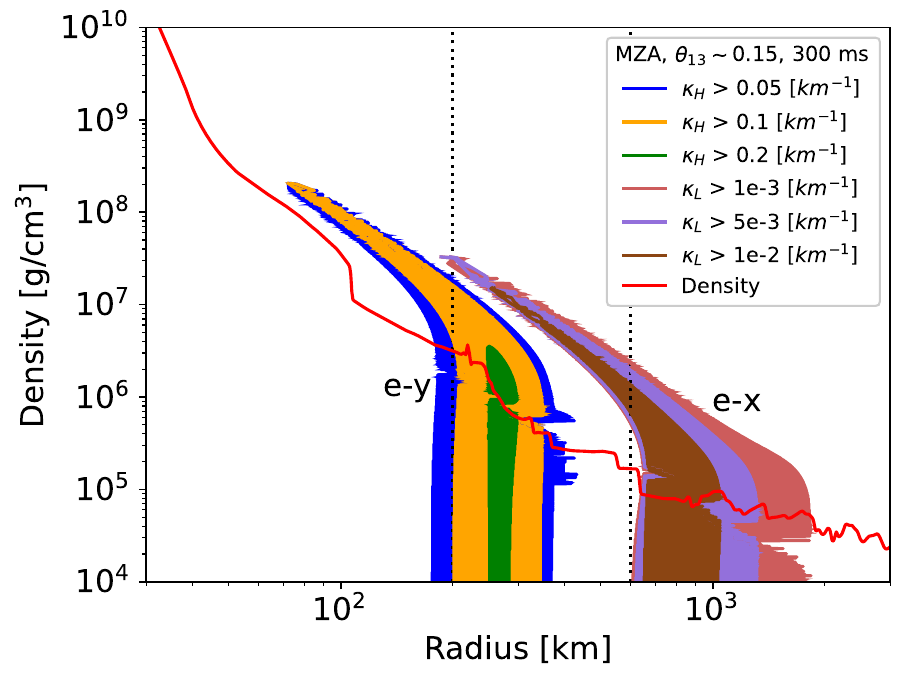}
    \includegraphics[width=0.9\linewidth]{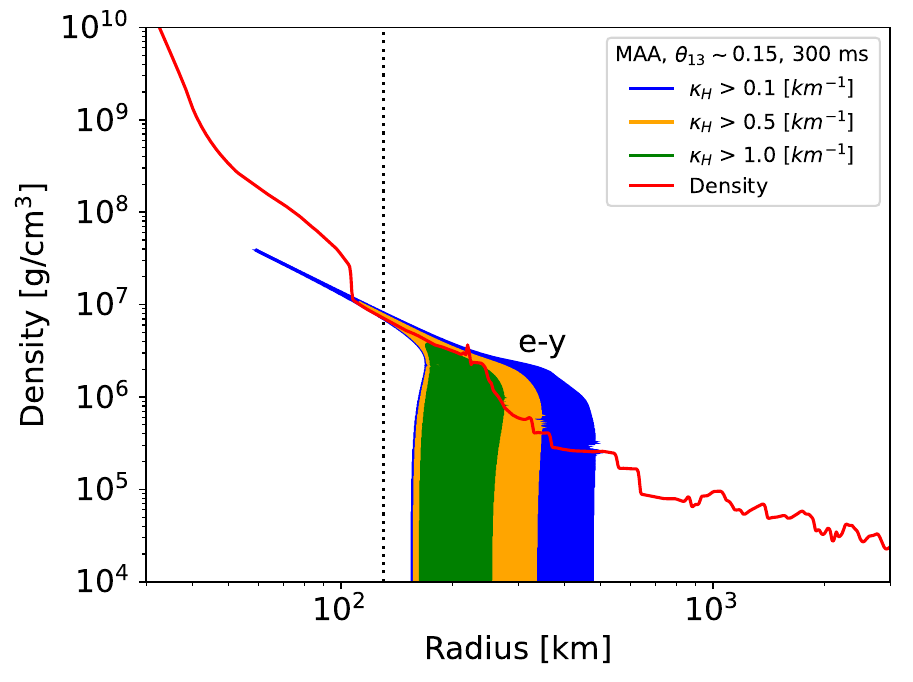}
    \includegraphics[width=0.9\linewidth]{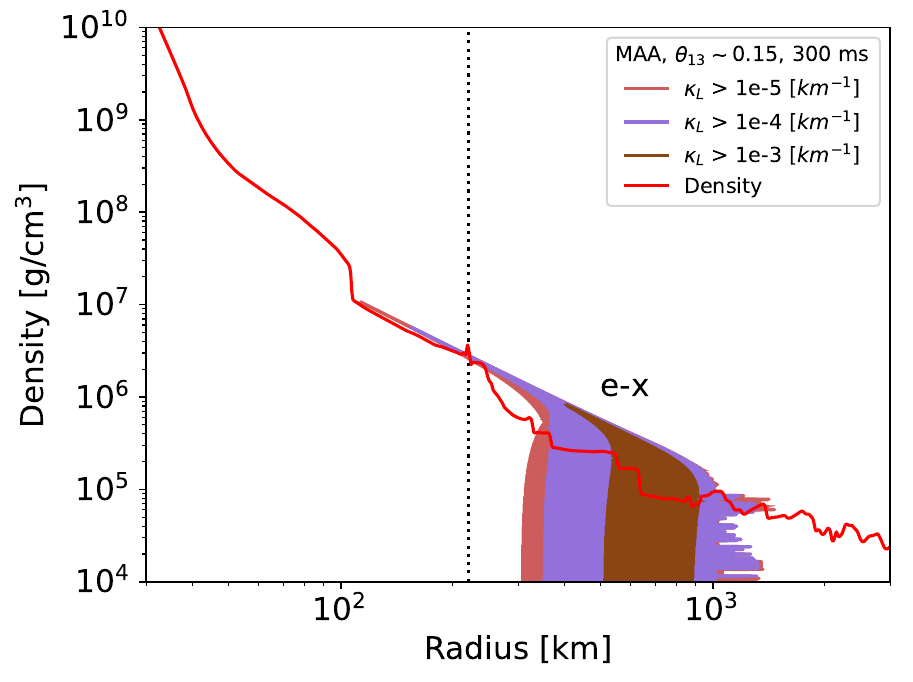}
	\caption{The same as Fig.~\ref{fig:foot_100_MZA}, but for $300\mathrm{~ms}$ and we also show both the MAA and MZA cases. 
	The top panel is the MZA case in both the $e-x$ and $e-y$ sectors.
	The middle and bottom panels are the MAA case in the $e-x$ and $e-y$ sectors, respectively.
	Here, we show only the non-small mixing angle case.
	}
	\label{fig:foot_300}
\end{figure}

Figure \ref{fig:foot_300} shows the contour maps in the MZA and MAA cases at $300\mathrm{~ms}$.
Again, the onset radius in numerical results (vertical dotted) is consistent with  the intersection points of the density profile (red solid) with the instability regions.
In the MZA case, the density profile invades into the instability regions and flavor conversions truly occur, different from the $100\mathrm{~ms}$ snapshot.
Multiple spectral crossings provide unstable modes and break the self-induced suppression at small radius.
In the MAA case, the density profile enters the instability region at $100\mathrm{~km}$ and passes through until larger radii. 
Therefore, the growth rate does not become weaker suddenly and flavor instability  sufficiently evolves, unlike in the $100\mathrm{~ms}$ snapshot.
Interestingly, comparing the MAA case with the MZA case, we find that the narrow region of footprint in the MZA case stays on higher density and flavor instability in the MAA case is weaker than that in the MZA case.
Previous work \cite{Chakraborty:2014} has reported that the MAA instability is easily suppressed for neutrino spectra with multiple crossings.
Actually, the lowness of the narrow region indicates suppression behaviors in the MAA case.
\\

\subsection{Signal prediction}
\label{Sec3c}
Finally, we make signal predictions at current and future detectors, namely, Super-Kamiokande and DUNE.
We numerically calculate the flavor evolution including vacuum, matter, and collective effects until $1500\mathrm{~km}$.
At this radius, the neutrino density is small enough and self-induced effects cease.
The emitted neutrinos propagate through the H- and L-resonance in the outer layer of the progenitor and undergo additional mixing.
Here, we assume these MSW effects produce adiabatic resonances and simple flavor mixing.
For NO case, the MSW flavor mixing is described in \cite{Dasgupta:2008b} as
\begin{eqnarray}
    f^{\mathrm{obs}}_{\nu_e} &&= s_{12}^2\left(s_{13}^2 f^{\mathrm{CNO}}_{\nu_e} + c_{13}^2 f^{\mathrm{CNO}}_{\nu_y}\right) + c_{12}^2 f^{\mathrm{CNO}}_{\nu_x} \\
    f^{\mathrm{obs}}_{\bar{\nu}_e} &&= c_{12}^2 \left(c_{13}^2 f^{\mathrm{CNO}}_{\bar{\nu}_e} + s_{13}^2 f^{\mathrm{CNO}}_{\bar{\nu}_y}\right) + s_{12}^2 f^{\mathrm{CNO}}_{\bar{\nu}_x},
\end{eqnarray}
where $f^{\mathrm{obs}}$ is a neutrino flux at the surface of SN and $f^{\mathrm{CNO}}$ is after collective neutrino oscillation vanishes at $1500\mathrm{~km}$.
We evaluate the event rate, assuming a SN of an $8.8 M_{\odot}$ progenitor at $d=10\mathrm{~kpc}$ as a representative distance in our Galaxy.

\begin{figure}[t]
	\centering
    \includegraphics[width=0.9\linewidth]{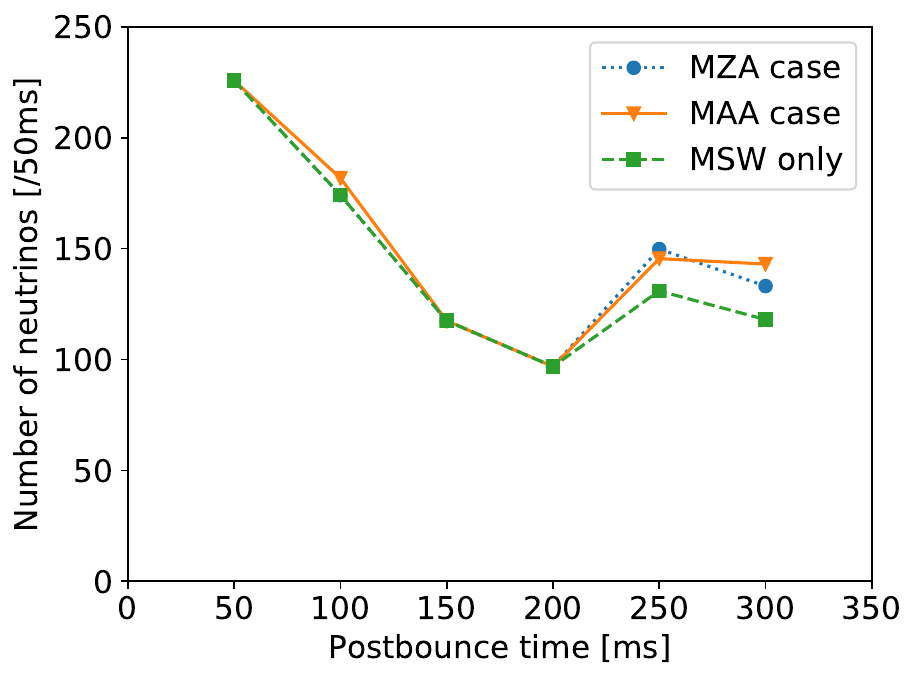}
	\caption{Detected IBD event rate per $50\mathrm{~ms}$ bins at Super-Kamiokande from an ECSN at $10\mathrm{~kpc}$.
	The MAA case, the MZA case, and the MSW only case are shown as orange solid, blue dotted, and green dashed lines, respectively.
	The inclusion of collective neutrino oscillation tends to increase the IBD event rate.
	}
	\label{fig:SK}
\end{figure}
\begin{figure}[t]
	\centering
    \includegraphics[width=0.9\linewidth]{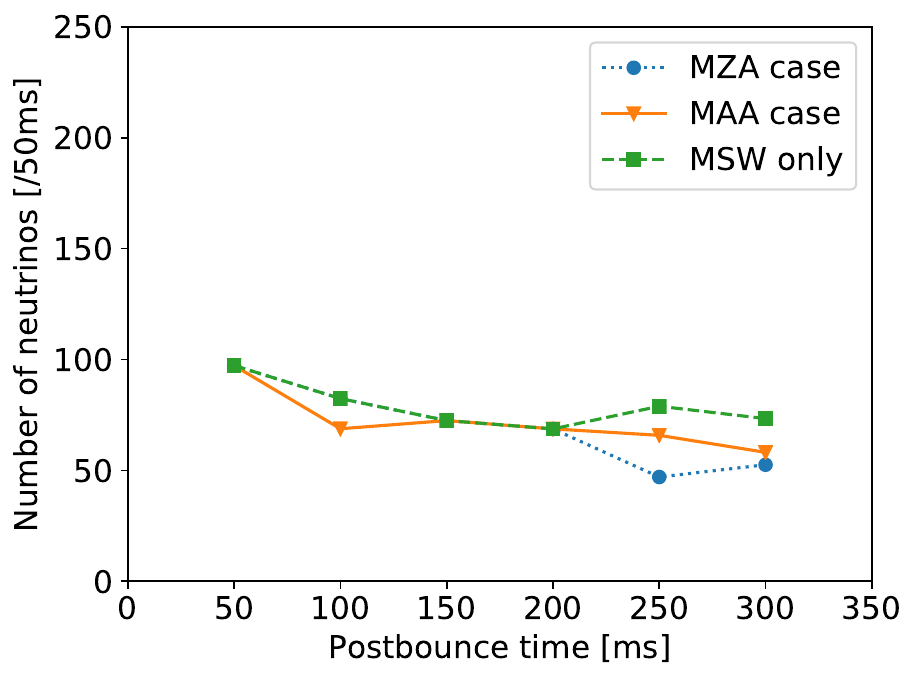}
	\caption{The same as Fig. \ref{fig:SK}, but for $\nu_e$ CC at DUNE.
	In contrast to $\bar{\nu}_e$, the inclusion of collective neutrino oscillation tends to decrease the $\nu_e$ event rate.
	}
	\label{fig:DUNE}
\end{figure}

Super-Kamiokande has high sensitivity to inverse beta decay (IBD) detecting electron anti-neutrinos.
The threshold energy of the recoil positron kinetic energy is $3.5\mathrm{~MeV}$ \cite{Sekiya:2013,Sekiya:2016}.
For $\bar{\nu}$ is $E_{\mathrm{th}} \sim 4.79\mathrm{~MeV}$ because of the mass difference between the proton and neutron.
We take the cross section $\sigma$ of IBD from Ref.~\cite{Strumia:2003} and the detector size to be the full inner-detector volume of $32.5\mathrm{~kton}$ in the observation of short burst events like SNe, larger than the fiducial volume \cite{Scholberg:2012}.
The event rate is simply evaluated as,
\begin{equation}
    \frac{\mathrm{d}N}{\mathrm{d}t} = \frac{N_{\mathrm{tar}}}{4\pi d^2}\int_{E_{\mathrm{th}}}\mathrm{d}E~ f_{\nu_{\alpha}}(E)\sigma(E),
\end{equation}
where $N_{\mathrm{tar}}$ is the number of target particles in the detector and $f_{\nu_{\alpha}}(E)$ is the angle-averaged neutrino flux.
In Super-Kamiokande, the target particle is protons of pure water $\mathrm{H_2 O}$ in the tank.
Here, we do not consider actual detection efficiency, detector responses, or energy resolution for simplicity.

Figure \ref{fig:SK} shows the time evolution of the event rate per $50\mathrm{~ms}$ at Super-Kamiokande.
Here, we present three types of cases: the MZA, MAA, and MSW only cases, all for NO.
The behaviors of suppression phenomena are switched as time passes and four phases emerge.
This phase evolution is due to the shock expansion.
The inclusion of collective neutrino oscillation tends to increase the event rate.
This is related to the high-energy tail of the neutrino spectrum.
The higher the neutrino energy, the larger the cross section of IBD, so the high-energy tail is important for the event number.
Since the high-energy tail is more prominent in the non-electron type neutrinos, how much the spectral tail of $\nu_X$ is mixed into electron neutrinos explains the difference in event rates.
In the NO case, the transition probability through the H-resonance is $\cos^2\theta_{13}\sim 0.98$ and that through the L-resonance is $\cos^2\theta_{12}\sim 0.7$.
In the MSW only case, only $30\%$ of the $\nu_x$ spectrum is mixed into electron neutrinos.
On the other hand, the inclusion of collective neutrino oscillation provides additional $e-X$ conversions and leads to increase of the event rate by $\sim 20\%$ in $300\mathrm{~ms}$ case.

The Deep Underground Neutrino Experiment (DUNE) is a liquid argon time-projection chamber which can detect electron neutrinos via the charged-current reaction, $^{40}\mathrm{Ar}+\nu_e \to e^{-} + ^{40}\mathrm{K}^*$, where the de-activation of K$^*$ can be picked up with a photon system.
Therefore, simultaneous observations with DUNE and Super-Kamiokande enables complementary information of ordinary and anti electron neutrinos.
DUNE will be composed of four detectors and the total fiducial volume of liquid argon is designed to be $40 \mathrm{~kton}$.
The threshold energy of this charged-current reaction has not been precisely determined yet \cite{Ankowski:2016}, and here we assume that the electron energy cut-off is $5 \mathrm{~MeV}$.
Due to the energy difference between $\mathrm{^{40}\mathrm{Ar}}$ and $^{40}\mathrm{K}^*$, the threshold energy for electron neutrinos is $8.28\mathrm{~MeV}$.
We take the cross section from Ref.~\cite{Suzuki:2013}.

Figure \ref{fig:DUNE} shows the time evolution of event rate at DUNE, again per 50 ms bins and a SN distance of 10 kpc.
As with the Super-Kamiokande case, the shock propagation affects the expected neutrino event rate.
However, the effect of collective neutrino oscillation in DUNE is the opposite.
The inclusion tends to decrease the event rate in DUNE.
This trend is derived from complete conversions through the H-resonance.
The tail components are important similar to Super-Kamiokande and the H-resonance switches electron neutrinos and non-electron neutrinos.
On the other hand, the final spectrum of electron neutrinos are not purely the non-electron type spectra because of collective flavor conversions that occur before the H-resonance in the MZA and MAA cases.
So the event rate also decreases because collective neutrino oscillation decreases the contribution of high energy tail of original non-electron type neutrinos on detected electron neutrino spectrum.
Also, the difference between the MZA and MAA cases is clearly shown.
At $300\mathrm{~ms}$, collective neutrino oscillation easily occurs because multiple crossings provide unstable conditions even in the MZA case.
On the other hand, the MAA instability causes multi-angle decoherence and leads to flavor equilibrium of the neutrino ensemble.
The observed electron neutrino spectrum has relatively high contribution of the spectral tail of the $\nu_X$ after passing through the MSW resonances because the spectral shapes of the $\nu_x$ and $\nu_y$ are identical in the MAA case.
Therefore, the electron neutrino spectrum in the MZA case is more different from in MSW only case than in the MAA case.
\\

\section{Conclusions}
\label{Sec4}
We have performed the first-ever numerical study of three-flavor collective neutrino oscillation considering three-dimensional momentum distribution in a realistic ECSN model of an $8.8 M_{\odot}$ progenitor.
The azimuthal angular distribution of emitted neutrinos triggers axial-symmetry breaking and causes MAA effects.
The MAA instability can be enhanced even in NO, under which axial symmetric (MZA) case is potentially stable.
To interpret our results, we have also extended the linear stability analysis into three-flavors including mixing angles.
Finally, we make signal predictions for a Galactic supernova event considering current and future detectors by using our obtained neutrino spectra.

We found that the MAA effects can occur during time snapshots when the MZA instability is completely suppressed.
The oscillation feature is similar to the flavor conversion induced by the bimodal instability in the IO case and spectral splits appear.
The $e-y$ split is dominant, but a tiny $e-x$ conversion also appears.
The growth rate in the $e-x$ sector is much smaller than in the $e-y$ sector and provides partial flavor conversion.
Also, interesting is the influence of the mixing angle $\theta_{13}$ on the $e-x$ sector evolution.
In the MZA case, the application of the latest non-small experimental value rises the instability associated with $\Delta m^2_{21}$ and can cause $e-x$ flavor conversions even in NO.
However, the addition of the MAA effect reverses the oscillation criterion and weakens the growth rate.
In this sense, the MAA case in NO resembles the bimodal instability in IO at early time snapshots.
Considering the axial-symmetry breaking, we can therefore find spectral split features in both IO and NO.

We found that the neutrino event rates at Super-Kamiokande and DUNE show systematic changes when collective neutrino oscillation survives against the multi-angle matter suppression.
During the neutronization burst and the accretion phase, collective neutrino oscillation is suppressed by the excess of electron neutrinos and heavy accreting matter; we call this the suppression phase.
Later, flavor conversions overcome matter suppression around $100\mathrm{~ms}$ in our ECSN model, and we call this the window phase.
Neutrinos undergo matter suppression effects again around $150\mathrm{~ms}$ as the matter profile increases and the neutrino flux decreases; we call this the re-suppression phase.
Finally, the self-induced effects revive around $300\mathrm{~ms}$ during the cooling phase; we call this the revival phase.
Note that these phases are only what we found on our ECSN model and they may not necessarily be generic.

In the MZA case, the re-suppression phase starts very early and the window phase does not exists.
Thus, the consideration of the MAA effect opens the possibility for the multi-angle matter and self-induced suppression in NO and provides a different evolution of the neutrino event rate.
During the cooling phase, small flavor asymmetry and multiple crossings break the synchronization of neutrino ensembles and make possible flavor conversions in both mass ordering.
The behavior of multi-angle decoherence in the MAA case changes the final spectral shapes and the difference in the neutrino event rate is enhanced.

In summary, we investigated the influence of the MAA instability on flavor evolution within the three-flavor framework.
The inclusion of the MAA effects alters the suppression behaviors and provides additional time evolution in the signal prediction.
However, the behaviors found here are for our employed ECSN model, and expected results would be different for other more massive progenitor cases.
In particular, the window phase could be shorter or vanish due to stronger matter suppression for iron-CCSN models.
Also, to obtain a treatable description, we imposed the assumption that the transverse evolution can be ignored in order to extend the bulb model.
The small-scale transverse variations should actually grow and could smear the observed signals at the Earth.
Also, the inclusion of the temporal evolution can compensate the matter-induced phase dispersion and enhance the impact of the inhomogeneous modes on the flavor evolution.
In addition, we also ignored the neutrino halo effect and fast flavor conversion.
These effects require the inclusion of neutrino scattering and more detailed supernova simulations.
But they are expected to influence the signal prediction.
The investigation of the global solutions including the multi-dimensional spatial and temporal evolution is still open and there is still much work needed to challenge the computational complexity of the self-induced interactions.
\\

\begin{acknowledgments}
M.Z. is supported by the Japan Society for Promotion of Science (JSPS) Grant-in-Aid for JSPS Fellows (No. 20J13631) from the Ministry of Education, Culture, Sports, Science, and Technology (MEXT).
S.H.\ is supported by the U.S.\ Department of Energy Office of Science under award number DE-SC0020262 and NSF Grants Nos.\ AST-1908960 and PHY-1914409. K.K. is supported by Research Institute of
Stellar Explosive Phenomena at Fukuoka University and the associated project (No. 207002). This work is supported by JSPS KAKENHI
Grant Number (
JP17H01130, 
JP17H06364, 
JP17H06357, 
JP18H01212, 
and JP20H05249
).
This research was also supported by MEXT as “Program for Promoting 
researches on the Supercomputer Fugaku” (Toward a unified view of 
the universe: from large scale structures to planets) and JICFuS.
Numerical computations were in part carried out on Cray XC50 at Center for Computational Astrophysics, National Astronomical Observatory of Japan.
\end{acknowledgments}

\nocite{*}
%

\end{document}